\documentclass[12pt,titlepage]{article}
\usepackage{latexsym}
\usepackage{amssymb}
\usepackage{graphicx}
\usepackage{color}
\usepackage{psfrag}
\usepackage{fancybox}
\usepackage{psfig}
\renewcommand{\to}{\rightarrow}

\newcommand{\beq}{\begin{equation}}
\newcommand{\eeq}{\end{equation}}
\newcommand{\bea}{\begin{eqnarray}}
\newcommand{\eea}{\end{eqnarray}}
\newcommand{\eq}{\begin{equation}}
\newcommand{\en}{\end{equation}}
\newcommand{\eqa}{\begin{eqnarray}}
\newcommand{\ena}{\end{eqnarray}}

\def\e{{\rm e}}

\def\tr{{\rm tr}}

\begin{document}
\thispagestyle{empty}
\begin{titlepage}
\addtolength{\baselineskip}{.7mm}
\thispagestyle{empty}
\begin{flushright}
\end{flushright}
\vspace{10mm}
\begin{center}
{\large 
{\bf Lectures on random matrix theory and symmetric spaces
}}\\[15mm]
{
\bf 
U.~Magnea 
} \\
\vspace{5mm}
{\it Department of Theoretical Physics, University of Torino \\
and INFN, Sezione di Torino\\
Via P. Giuria 1, I-10125 Torino, Italy} \\ 
blom@to.infn.it
\\[6mm]
\vspace{13mm}
{\bf Abstract}\\[5mm]
\end{center}

In these lectures we discuss some elementary concepts in connection with
the theory of symmetric spaces applied to ensembles of random matrices. We
review how the relationship between random matrix theory and symmetric spaces 
can be used in some physical contexts. 

\end{titlepage}

\newpage
\setcounter{footnote}{0}
\tableofcontents

\section{Introduction}

\subsection{Outline of the lectures}
\label{sec-outline}

In these lectures, that have been elaborated in part from the review
by the author and Caselle \cite{SS}, we will deal with some elementary
concepts relating to the description of random matrix ensembles as
symmetric coset spaces.  Some figures have been added to the text as
illustrations. These have been borrowed from various reviews and
papers on random matrix theory with permission from the authors and
publishers. 

As the conference organizers asked me to give {\it elementary}
lectures, these lecture notes should be accessible to a wide audience.
Even though none of the material presented here is entirely new,
perhaps these lectures can still serve the purpose of stimulating
interest in the research on random matrices.

In the first lecture we will introduce some of the physical systems
where Random Matrix Theory (RMT) is a useful tool. Then we will give a
brief overview of the most important elements in random matrix
theory. Part of the material here is based on excerpts from the
excellent review by Guhr, M\"uller--Groeling and Weidenm\"uller
\cite{GMW}. Since the audience consists of mathematicians, most of
whom are not working in this field, we will try to be as clear as
possible in describing what RMT is used for, why it is relevant, and
how its predictions are compared to experimentally or numerically
measured spectral fluctuations. In particular we will discuss the
concepts of unfolding and universality.

The second and third lectures are essentially a shortened version of
some of the material presented in \cite{SS}, with some modifications
and additions.  In the second lecture we will see that hermitean
random matrix ensembles can be identified with {\it symmetric coset
spaces} related to a compact symmetric subgroup. This was realized
early on by Dyson \cite{Dyson,DysonSS} and by H\"uffmann \cite{Huff}, but
the advantages due to this fact for a long time was not frequently
understood by physicists in applications to the above--mentioned
systems. The well--known theory of simple Lie algebras and symmetric
spaces \cite{Helgason,Helgason2,Gilmore} can be applied in random
matrix theory to obtain new results in various physical contexts
\cite{SS}. Key elements are the classification of symmetric spaces in
terms of root systems and the theory of invariant operators on the
symmetric manifolds.

We will discuss elements from the theory of symmetric spaces, using
explicit examples drawn from low--dimensional Lie algebras. As we will
see, symmetric spaces of positive, zero, and negative curvature
correspond to well--defined types of random matrix ensembles.  Along
the way, we will also discuss the identification of various elements
of RMT with the corresponding quantity on the symmetric space. In
particular it will be clear that eigenvalue correlations in random 
matrix theory has a geometric origin in the root systems characterizing
the symmetric space manifolds.

In the third and last lecture we discuss a few examples of
applications of symmetric coset spaces in random matrix theory. A few
more topics have been discussed in \cite{SS}. As the first and most
important application, we discuss the classification of disordered
systems arising from the Cartan classification of symmetric coset
spaces. Most of the hermitean random matrix ensembles corresponding to
symmetric spaces have known physical applications. We will not have
time to discuss these here, but physical applications of the ensembles
appearing in Table~\ref{tab3a}, along with more random matrix
ensembles, were discussed or at least mentioned in \cite{SS}, where
references to the literature can also be found. (I apologize to those
authors whose work has been overlooked here, as I am certain not to be
aware of all the published work on applications of random matrix
ensembles.)

Our second example is the solution of the DMPK equation using the
theory of zonal spherical functions (the eigenfunctions of the radial
part of the Laplace--Beltrami operator on the symmetric space). The
DMPK equation is the scaling equation determining the probability
distribution of the transmission eigenvalues for a quantum wire as a
function of the length of the wire. It will be identified with the
equation of free diffusion on the symmetric space.

The last example we give is more weakly related to RMT (it is related
only through the diffusion equation) and concerns the integrability of
certain $1d$ models referred to as Calogero--Sutherland models.  There
is a connection between these models and the theory of Lie algebras:
Calogero--Sutherland models are closely related to root systems
of Lie algebras or symmetric spaces.  Olshanetsky and Perelomov
\cite{OlshPere} provide an exact statement as to when these models are
integrable and directly express the physical integrals of motion as
the algebra of Laplace operators related to the Lie algebra or
symmetric space. These results lead to a detailed list of spectra and
wave functions for a variety of quantum systems.

\subsection{A few comments on the classification scheme}

It is interesting to note that a wide variety of microscopically
different physical systems can be described by the same type of
spectral fluctuations. RMT describes the generic features of the
spectrum, without regard to dynamical principles or details of the
interactions. This allows a separation of generic spectral
fluctuations from system--specific ones. The only input in RMT is
symmetry through the postulated invariance of the various ensembles.
This scenario leads to a classification of physical disordered systems
into symmetry classes characterized by universal spectral
behavior.

In the context of the classification scheme a new paper by Heinzner,
Huckleberry and Zirnbauer \cite{HHZ} should be mentioned, even though
none of the material there will be covered in these lectures. In this
paper, the authors prove the correspondence between symmetric spaces
and symmetry classes of disordered fermion systems with quadratic
Hamiltonians.  This is done by considering both unitary and
antiunitary symmetries and then removing the unitary symmetries by
considering the decomposition of the space of good Hamiltonians into
blocks associated with unitary subrepresentations in the Nambu space
of fermionic field operators. The relevant structures on this space
are transferred to a space of homomorphisms where the unitary symmetry
group acts trivially. The various cases occurring for the remaining
anti--unitary symmetries then lead to the classification in the
symmetric space picture.  The authors also observe that when second
quantization is undone (as it is in the physical systems corresponding
to the new symmetry classes arising in physics in addition to Dyson's
symmetry classes), a remnant of the canonical anticommutation
relations of the fermionic field operators is imposed on the Nambu
space. This is the reason why we get new structures in the physical
context of disordered fermions.

\subsection{Topics outside the scope of these lectures}

Of course, a significant fraction of the (vast) literature on the
theory and applications of RMT will {\it not} be covered here.  This
applies also to various types of extensions of simple hermitean random
matrix models, multimatrix models, and several phenomena in random
matrix theory, e.g. parametric correlations and phase transitions, and
theoretical issues like universality proofs and the supersymmetric
formalism.  The obvious reason is that our focus should be on the
relationship with symmetric coset manifolds. A more extensive
introduction to random matrix theory as well as experiments and
numerical simulations performed to disclose RMT behavior in spectra
can be found in \cite{GMW}, that gives a good overview of the various
aspects of random matrix theory.

Note that even the list of simple hermitean random matrix ensembles
mentioned in these lectures is not complete, though we do discuss the
main ones.  A few more ensembles were briefly discussed in
\cite{SS}. In addition there is a large number of {\it non--hermitean}
random matrix ensembles. There has been some recent activity in this
field, where most of the papers have dealt with the problem of finding
the eigenvalue distribution in the complex plane. The concept of
orthogonal polynomials has also been extended to the non--hermitean
case (see e.g. \cite{cop}).

In the applications discussed here, random matrices are used to
describe statistical fluctuations in the spectra of quantum operators.
In some field theoretical contexts, random matrices are employed in a
different way. In quantum gravity, they describe random discretized
surfaces corresponding to string world sheets of different genus. The
partition function corresponding to an integral over the gravitational
field is the sum of such surfaces, much like in the large $N$
expansion in QCD due to 't Hooft. In such a context, it is the field
rather than a quantum operator that is substituted by a random
matrix. We will not discuss these applications of random matrices. For
an introduction see \cite{AmQ}.

\section{Chaotic systems and random matrices}
\label{sec-chao}
\setcounter{equation}{0}

The wide range of physical problems where random matrix theory
can be successfully applied has made it into an important branch of
mathematical physics. As an introduction to the topic, we will give a brief
historical survey of the developments which led to the present
situation. 

\subsection{Many--body systems and Wigner--Dyson ensembles}

\begin{figure}
\centerline{
\psfig{file=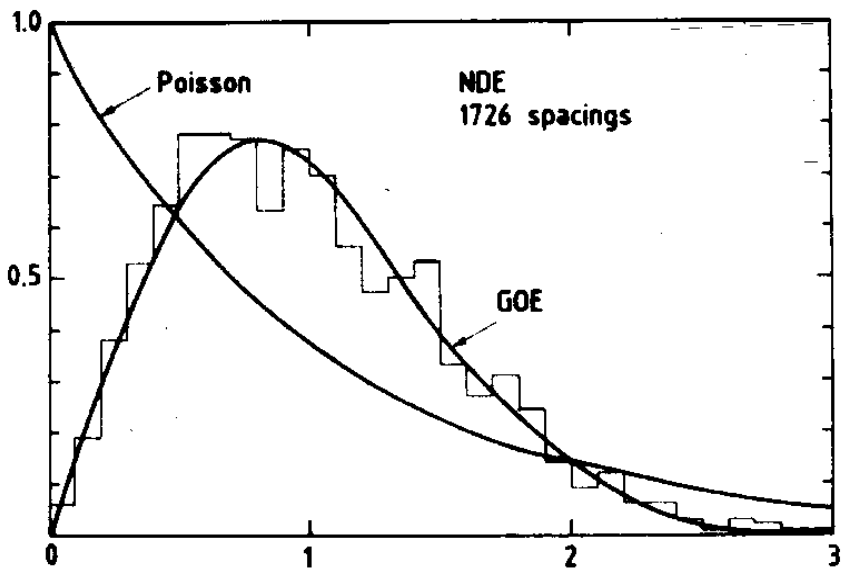,width=4.5in}
}
\caption{
Nearest neighbor spacing distribution for the ``Nuclear Data
Ensemble''  comprising 1726 spacings (histogram) versus $s=S/D$ with
$D$ the mean level spacing and $S$ the actual spacing. Note the level 
repulsion in random matrix theory (GOE denotes the Gaussian Orthogonal 
random matrix Ensemble) at zero spacing as compared to the Poisson 
distribution. Taken from reference \cite{bohigas}. Used with permission.} 
\label{fig1}
\end{figure}

In the 1950's, Wigner \cite{Wigner} developed a theory of random
matrices to deal with resonance spectra of heavy nuclei. Experiments
with neutron and proton scattering gave precise information on levels
far above the ground state, whereas nuclear structure models could
only predict the positions of levels close to the ground state. Wigner
conceived of a new way of studying the spectrum: a statistical theory
that could not predict individual energy levels, but described, in
Dyson's words, ``the general appearance and the degree of irregularity
of the level structure'' \cite{Dyson}. This provided a tool for
studying complex spectra.  Wigner's theory dealt with ensembles of
random matrices modelling the Hamiltonians of nuclei. In the early
1960's, Dyson developed Wigner's approach further in a series of papers
\cite{Dyson} where he treated scattering matrix ensembles. Typical of
the spectra obeying Wigner--Dyson statistics is that energy levels are
correlated and repel each other.  Such a level repulsion is not
present in spectra obeying Poisson statistics (cf. Fig.~\ref{fig1}).

\vskip1cm

Shortly thereafter RMT was applied to the spectra of atoms. Later on,
modern laser spectroscopy has allowed a comparison with the complex
spectra of polyatomic molecules. Nuclei, atoms and molecules are all
examples of {\it complex many--body systems} with a large number of
degrees of freedom.

\subsection{Quantum chaos}

\begin{figure}
\centerline{
\psfig{file=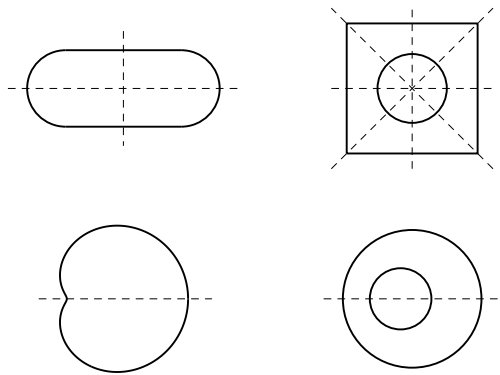,width=3in,angle=-90}
}
\caption{
  Four important billiards: the Bunimovich stadium
  (top left), the Sinai billiard (top right), the Pascalian
  snail (bottom left) and the annular billiard (bottom right).
  Reprinted from reference \cite{GMW} with permission from
  authors and from Elsevier. 
}
\label{figch11}
\end{figure}

It was later realized that the Wigner--Dyson ensembles could be
applied to the description of {\it chaotic quantum systems}.  In
these, a quantum particle is reflected elastically at the boundaries
of some given domain, so that the total energy is constant. In such a
system, all constants of the motion (except the energy) are destroyed
by randomness, and there are no stable periodic orbits.  The system is
referred to as a {\it quantum billiard} (see Fig.~\ref{figch11}).  Due
to the shape of the domain, normally chosen to be two--dimensional,
the trajectory of the particle is completely random. Such a system may
have just a few degrees of freedom and may be simulated by microwave
cavities. The reason is that the wave equation for the electromagnetic
field in the cavity has the same form as the Schr\"odinger equation
for a two--dimensional quantum billiard, if the geometry of the cavity
and the boundary conditions are chosen appropriately. In this
Schr\"odinger equation, the Hamiltonian for the free particle is
simply the Laplace operator.  With appropriate boundary conditions the
system may also be equivalent to a vibrating membrane.

Spectra of such systems with up to a thousand eigenmodes have been
studied experimentally (for a list of references, see \cite{GMW}).  The
results can be interpreted in terms of classical chaos. The general
picture emerging from such experiments is that if the shape of the
cavity is such that the classical motion in it is integrable
(regular), the spectrum behaves according to a Poisson
distribution. If the corresponding classical motion is chaotic, the
spectral fluctuations behave in accordance with the Wigner--Dyson
ensembles. This is also the content of the famous (but unproven) {\it
Bohigas conjecture}.

\subsection{Mesoscopic systems, BdG and transfer matrix ensembles}

A newer class of interesting quantum systems with randomness is
provided by {\it disordered metals}, i.e.  micrometer--sized metal
grains, of which we wish to study the transport properties when the
grain is connected to electron reservoirs through (ideal) leads (for a
review, see \cite{Beenakker}). Such microstructures can be fabricated in a
cavity in a semiconductor. They can be made so thin that the electron
gas inside them is effectively two--dimensional. Such structures are
quantum systems, yet they are large enough for a statistical
description to be meaningful. Therefore they are referred to as {\it
mesoscopic systems}. The {\it transmission eigenvalues} related to the
transfer matrix formalism determine the conductance, which is the main
observable. We will discuss these systems more explicitly in the third
lecture.

Disorder in such systems arises because of randomly distributed
impurities in the metal. The electrons in the conduction band are
scattered elastically by the resulting random potential as we apply a
voltage drop across the metal sample. Such a system is called {\it
diffusive}. At low temperature (below 1K), inelastic electron--phonon
scattering, which changes the phases of the electrons in a random way,
can be neglected. The electrons are therefore phase coherent over the
length of the sample.  In the diffusive regime (with the conductance
decreasing linearly with sample size) we expect Wigner--Dyson
statistics to apply to the spectrum of energy levels (at least up to
an energy separation given by the so--called Thouless energy).  The
random matrix theory of quantum transport, however, deals mostly with
the transmission eigenvalues. It relates the universality of transport
properties (i.e., the independence of these of sample size, degree of
disorder, etc.) to the universality (to be discussed) of correlation
functions in random matrix theory.

The behavior of the conductance depends critically on the
dimensionality of the sample. In $1d$ all states are localized and the
sample is an insulator, while $d=2$ is the critical dimension and
delocalization occurs for $d>2$.  If the sample has the shape of a
grain, it is called a {\it quantum dot}.  A {\it quantum wire} is a
very thin (quasi--one dimensional) metal wire with similar properties,
that allows to study the scaling properties of observables related to
transport as a function of the length of the wire through a
generalized diffusion equation. Quasi--$1d$ wires are particularly
good laboratories for testing RMT. Since they are not strictly
one--dimensional, they possess a diffusive regime (cf. the comment
above on dimensionality); still, non--perturbative analytical methods
are applicable (scaling equation, field theoretical methods).

The phenomenon of {\it localization} occurs when the previously
extended Bloch wave functions of the multiply scattered electrons are
cancelled by destructive interference. This happens when the density
of impurities reaches a critical value. As a consequence, the metal
sample goes from being a conductor to being an insulator. In the
localized (insulating) regime the conductance decays exponentially as
a function of sample size and the typical length scale of the decay of
electron wave functions is given by the localization length $\xi $.  In
the localized regime, wave functions of nearly degenerate states may
have a very small overlap. As a consequence, the level repulsion
typical of the Wigner--Dyson ensembles is suppressed and we expect
Poisson statistics (uncorrelated eigenvalues) on length scales larger
than $\xi $.

\begin{figure}
\centerline{
\psfig{file=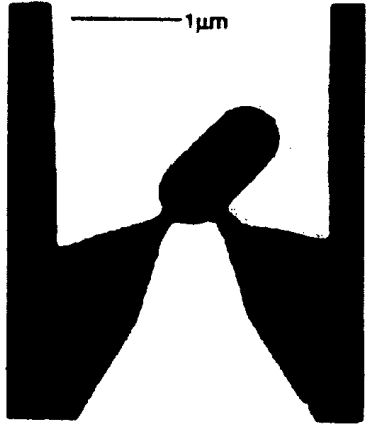,width=2in}
}
\caption{ Electron micrograph of a quantum dot in the shape of a
  Bunimovich stadium, with $1~\mu$m bar for scale. The electrons can
  move in the black area. Two leads are coupled to the stadium.
  Reprinted with permission from ref. \cite{marcus}. Copyright 2005 by
  the American Physical Society.}
\label{figch10}
\end{figure}

We may also consider a metal grain in which electron
scattering takes place at the boundaries. In the latter case the mean
free path of the electrons is large compared to the size of the system
and we speak of a {\it ballistic} system. A ballistic quantum dot is
very similar to a billiard and is described by Wigner--Dyson statistics
(see Fig.~\ref{figch10}).

In addition to these systems, heterostructures consisting of
superconductors in conjunction with normal metals are successful
candidates for a random matrix theory description. In these we have
particle--hole symmetry and the (first--quantized) Hamiltonians in
this case are described by four new so--called Bogoliubov--de Gennes
(BdG) ensembles \cite{AltZ}.

The existence of several mesoscopic regimes (localized, diffusive,
ballistic, with varying degree of disorder) enriches the phenomenology
and extends the applications of RMT beyond the ones discussed for
many--body and chaotic systems. In particular, the case of
normal--superconducting quantum dots leads to four entirely new
symmetry classes of Hamiltonians. 

In mesoscopic systems, the operator that is modelled by random matrix
ensembles is either the Hamiltonian or the transfer matrix. In the
random Hamiltonian approach, ensemble averages are calculated in the
supersymmetric formalism \cite{Bee154-155} using scattering theory.
(The supersymmetric formalism is partly discussed also in
\cite{GMW}. In addition see \cite{Vsusy} for QCD--related
applications.)  In the random transfer matrix approach, which we will
discuss more explicitly later, the transmission eigenvalues determine
important transport properties, of which the most central one is the
conductance determined in the Landauer--Lee--Fisher theory.  To every
ensemble of Hamiltonians corresponds an ensemble of transfer matrices
(and one of scattering matrices as well) describing the same physical
system.

\subsection{Field theory and chiral ensembles}

Random matrices are also useful in relativistic {\it quantum field
theory}.  In a gauge field theory, the interactions between the gauge
field and the fermions is expressed as a gauge field integration over
the fermion determinant, obtained by integration over the fermionic
(Grassmann) variables in the partition function.  The fermion
determinant involves the Dirac operator and depends on the gauge
fields. This operator can also be modelled by an ensemble of random
matrices. If chiral symmetry is present, these ensembles are called
chiral random matrix ensembles. The matrices in these ensembles have a
block--structure similar to the Dirac operator in the chiral basis,
and may possess zero modes. This application of random matrix theory
was developed by Verbaarschot et al. in the 1990's (for some early work, 
see \cite{earlyV}).

The procedure of substituting the Dirac operator with a random matrix
makes it possible to perform the integration over the gauge fields by
integrating over the ensemble of random matrices. By using techniques
borrowed from the supersymmetric formalism originally developed for
scattering problems in the physics of quantum transport, one can then
obtain the effective low--energy partition function in the gauge
theory \cite{effL}.  In the presence of spontaneous symmetry breaking,
it is expressed as an integral over the Goldstone manifold (i.e. over
the degrees of freedom associated with spontaneous symmetry breaking
in the gauge theory). This way the symmetry breaking pattern is also
obtained and in addition, sum rules for the eigenvalues of the Dirac
operator can be derived. Recently, there have been interesting new
developments in this field, relating the QCD partition function to
integrable hierachies (for a comprehensive review with references,
see the lecture series by Verbaarschot \cite{integrability}).

The order parameter for the spontaneously broken phase (the quark
condensate in QCD) is proportional to the density of Dirac eigenvalues
at the origin of the spectrum (i.e., at zero eigenvalue).  This is
expressed by the Banks--Casher formula.  Therefore the Dirac spectrum,
and especially the regime around the origin, is of major interest. Its
generic features can be studied in RMT. A wealth of numerical
simulations \cite{numerical} in lattice gauge theory show that data
agree with RMT predictions for level correlators and low--energy sum
rules for the Dirac eigenvalues. This shows that fluctuations of Dirac
eigenvalues are generic and independent of details of the gauge field
interactions. Two reviews with references to original work are given in
\cite{revVerb}.

Note that since these lectures are introductory, we will limit the
discussion to hermitean random matrix models. Therefore we will not
discuss the new developments in this field involving non--hermitean
random matrices (see e.g. \cite{newV}), even though these are very
intriguing as they allow a study of the case of nonzero baryon
chemical potential. The same is true for the new research developments
in non--hermitean random matrix theory in other branches of physics.
Such applications include for instance the description of
non-equilibrium processes, quantum chaotic systems with an imaginary
vector potential (related to superconductivity), and neural networks.

\section{What is random matrix theory?}
\label{sec-rmt}
\setcounter{equation}{0}

We have established that some of the things that may make a system
chaotic are: complex many--body or gauge field interactions, a random
impurity potential, or irregular shape (where irregular means any
shape leading to classical chaotic motion).  Because of the high
number of degrees of freedom or the complexity of the interactions in
chaotic systems, the quantum mechanical operators whose spectra and
eigenfunctions we are interested in may be unknown or inaccessible to
direct calculation. Such an operator may be a Hamiltonian, a
scattering or transfer matrix, or a Dirac operator, as we have seen
above.  These operators can be represented quantum mechanically by
matrices acting on a Hilbert space of states.
                                                                              
In studying nuclear resonances, Wigner \cite{Wigner} proposed
replacing the Hamiltonian of the nucleus by a random matrix taken from
some well--defined ensemble. How do we determine which ensemble is
appropriate? Wigner proposed choosing it so that the general {\it
global symmetries} of the underlying physical problem are preserved.
This means that we should choose random matrices with the same global
symmetries as the physical operator we are modelling. In doing so, we
will preserve the spectral properties that depend only on these
symmetries.  These are the aspects of the spectrum we will be able to
study using RMT.

As it turns out, the gross features of the random matrix theory
spectrum, like for instance the macroscopic shape of the eigenvalue
density, is usually not of interest for studying the physical system,
as it does not even remotely resemble the actual spectrum and is
non--universal in that it depends on the choice of random matrix
potential. The strength of the RMT description lies in the {\it
universal behavior of the eigenvalue correlators in a certain scaling
limit}. As we will see, this leads to universal spectral fluctuations
that are reproduced by a multitude of physical systems in the real
world, several of which we have discussed in the previous paragraph.

The statistical theory of Wigner \cite{Wigner} and Dyson \cite{Dyson}
is a theory of {\it random matrices}. These are matrices with random
elements chosen from some given distribution. In the applications to
be discussed here, we use them to represent a quantum mechanical
operator, whose eigenvalue spectrum and eigenstates are of physical
interest. In the classical example of the heavy nucleus, the
Schr\"odinger equation

\beq
H\Psi_i = E_i\Psi_i
\eeq

gives the energy spectrum of the nucleus with Hamiltonian $H$.  In
case this Hamiltonian is unknown or too complicated, we may choose to
study the eigenvalues of an appropriate ensemble of random matrices
instead.  The ensembles studied by Dyson in the early sixties were
ensembles of scattering matrices, but the principle is the same.

The random matrix eigenvalues should behave much like the energy
levels of the nucleus, if we choose our ensemble of random matrices
appropriately. To achieve this we analyze the {\it physical
symmetries} of the Hamiltonian and choose an ensemble of random
matrices with these same symmetries. The characteristics of the
spectrum that depend only on these symmetries will be present also in
the spectrum of the random matrix eigenvalues. These are the {\it
universal} features of the system, and they are obtained after removal
of the non--universal behavior through a proper rescaling
procedure. They do not depend on details of the physical interactions
or, in the RMT, on the choice of RMT potential.

The number of fields where random matrix models are applied nowadays is
growing fast. To convey the main ideas, we will concentrate on a few
major applications.

\section{Choosing an ensemble}
\label{sec-symmetries}
\setcounter{equation}{0}

What are the physical symmetries that determine the ensemble of random
matrices? Dyson's analysis shows that for the Hamiltonian ensembles
there are three generic cases:

1) The system is invariant under time--reversal (TR) and the total
spin is integer (even--spin case), {\it or} the system is invariant
under time--reversal and space--rotation (SR) with no regard to the
spin.  \footnote{Time reversal invariance arises from the fact that
(in the absence of magnetic fields), if $\Psi({\bf r},t)$ is a
solution of the Schr\"odinger equation, so is $\Psi^*({\bf r},-t)$. We
define the action of the (anti--unitary) time-reversal operator $T$ on
a state $\Psi$ by $T\Psi=K\Psi^*$, where $K$ is a unitary
operator. $T$ should reverse the sign of spin and total angular
momentum. This requirement can be satisfied by the choice $K={\rm
e}^{i\pi S_y}$ for spin rotation or $K={\rm e}^{i\pi J_y}$ for space
rotation, with a standard representation of spin or space rotation
matrices. Now $T^2=KK^*={\rm e}^{2i\pi S_y}=(-)^n=\pm 1$. These two
cases correspond to integer and half--odd integer spin (or presence or
absence of space rotation invariance in case of $J_y$). In the former
case, $K$ can be brought to unity by a unitary transformation $\Psi
\to U\Psi$ of the states, during which $K$ transforms into
$UKU^T$. Once this choice has been made, the only transformations on
$H$ and $\Psi$ allowed are $H\to RHR^{-1}$, $\Psi\to R\Psi$ with $R$
an orthogonal matrix.  In the latter case, a block--diagonal form for
$K$ may be chosen and only symplectic transformations are allowed on
$H$ and $\Psi $. In case there is no time reversal invariance, unitary
transformations on $H$ and $\Psi$ are allowed.}

2) The system is invariant under time--reversal but with no rotational
symmetry and the total spin is half--odd integer (odd--spin case).

3) The system is not time--reversal invariant. 

In all these cases, the ensemble of random matrices is invariant under
the automorphism

\beq 
H\to UHU^{-1}
\eeq

where $U$ is an orthogonal (case 1), unitary (case 3) or symplectic
(case 2) matrix.  It is this property that gives rise to the
identification of the manifolds of random matrices with symmetric
spaces. This identification is very useful, since it leads to the
possibility of applying the known theory of symmetric spaces in
physical contexts.

The properties of these ensembles are summarized in 
Table~\ref{tab0}.\footnote{The dual $Q^R$ of a matrix $Q$ consisting of
$N\times N$ real quaternions $q=q_0+{\bf q}\cdot {\bf \tau}$ is
$Q^R_{ij}=\bar{q}_{ji}$ where $\bar{q}=q_0-{\bf q}\cdot {\bf \tau}$
and $\tau_i=-i\sigma_i$ where $\sigma_i$ are Pauli
matrices. Self--dual means $Q=Q^R$.}

\begin{table}[ht]

\caption{Dyson's three universality classes of random matrices. The Dyson
index $\beta $ counts the number of degrees of freedom of the matrix
elements (each matrix element is a real, complex or quaternion number)
and is used to specify the ensemble. The ensembles are named the
orthogonal, unitary or symplectic ensemble, respectively, according to
the subgroup of invariance.
\label{tab0}}
\vskip5mm

\hskip35mm
\begin{tabular}{|c|c|c|c|c|}
\hline
$\beta$ & TR & SR   & $H$                        & $U$ \\
\hline
1       & X  & X    & real, orthogonal           & orthogonal \\
2       &    & (X)  & complex, hermitean         & unitary \\
4       & X  &      & real quaternion, self-dual & symplectic \\
\hline
\end{tabular}
\end{table}

The relevant physical symmetries depend on the system under
consideration.  Novel ensembles not included in this classification
arise if we impose additional symmetries or constraints.  This is the
case for the four universality classes of the Hamiltonians of
normal--superconducting (NS) quantum dots (for a detailed discussion
see \cite{AltZ}). Also here the four universality classes are
distinguished by the presence or absence of TR and SR, but with the
difference that the Hamiltonian possesses so--called particle--hole
symmetry. As electrons tunnel through the potential barrier at the NS
interface, a Cooper pair is added to, or subtracted from, the
superconducting condensate. Equivalently, an electron incident on the
interface is reflected as a hole (a phenomenon referred to as Andreev
reflection).

Three more chiral symmetry classes are realized in chiral gauge
theories. In this case the ensemble is chosen such that the chiral
symmetry $[\gamma_5,i\!\not{\!\!  D}]=0$, the zero modes and possible
anti--unitary symmetries $[Q,i\!\not{\!\! D}]=0$ of the Euclidean
Dirac operator (where $Q$ is some anti--unitary operator) are
reproduced by the ensemble. The properties of $Q$ determine if there
is a basis in which the Dirac operator is real or quaternion real. In
this case we also have three symmetry classes distinguished by the
index $\beta $.

As we will see later, to every hamiltonian ensemble there corresponds
a scattering matrix ensemble and a transfer matrix ensemble. The
hamiltonian and scattering matrix ensembles of a given physical system
are related to each other in a simple way: the hamiltonian ensemble is
simply the algebra subspace corresponding to the compact symmetric
space of the scattering matrix.  Because of the way transfer matrices
are defined, the ensemble of the transfer matrix of the same system is
not the corresponding non--compact space. Nevertheless, it will neatly
fit into the classification scheme. For scattering and transfer
matrices in the theory of quantum transport we use ensembles
constrained by the physical requirements of flux conservation,
time--reversal symmetry, and spin--rotation. Flux conservation imposes
the condition of unitarity on the scattering matrix, and determines
the symmetry class of the transfer matrix too. We will discuss some
explicit examples in the third lecture.

\section{General definition of a matrix model}
\label{sec-defrmt}
\setcounter{equation}{0}

Since many of the interesting quantum operators are physical
observables, we often deal with hermitean random
matrices. Non--hermitean operators are also of interest. The number of
fields where they are important is growing, and so is the research
effort devoted to this large class of random matrix
theories\footnote{Non--hermitean quantum operators are applied in
non--equilibrium processes and dissipative systems. Examples are
quantum chaotic scattering in {\it open} systems, conductors with
non--hermitean quantum mechanics, and systems in classical statistical
mechanics with non--hermitean Hamiltonians. They are also useful in
schematic random matrix models of the QCD vacuum at non--zero
temperature and/or large baryon number, where the fermion determinant
becomes complex.}. Here we will only be concerned with hermitean
random matrices.
 
A hermitean matrix model is defined by a partition function

\beq
\label{eq:partition}
Z\sim \int dH\, P(H)
\eeq

where $H$ is a square $N\times N$ hermitean matrix with random
elements $H_{ij}$ chosen from some given distribution. At the end of
our calculation, we will take the limit $N\to \infty $ to counteract
the fact that we are using, for technical reasons, a truncated Hilbert
space.  In equation~(\ref{eq:partition}) $P(H)$ is a probability
distribution in the space of random matrices and $dH$ is an invariant
measure in this space. Such a measure is required to give physical
meaning to the concept of probability: $P(H)dH$ is the probability
that a quantum operator in the ensemble will belong to the volume
element $dH$ in the neighborhood of $H$. Since the ensemble of
matrices $\{H\}$ does not in general form a group, the existence of
such a measure is not entirely trivial. Dyson \cite{Dyson} showed
that a unique invariant (Haar) measure $dH$ exists for the three
ensembles discussed in the previous section.

One can show that a probability distribution of the form

\beq
P(H)\sim {\rm e}^{-c\, {\rm tr} V(H)}
\eeq

is appropriate (see \cite{Mehta} for more details on this point).
Note that such a weight function is needed to keep the integrals from
diverging. Here $c$ is a constant and $V(H)$ is a matrix potential,
typically a polynomial with a finite number of terms. The simplest
choice is a quadratic potential, in which case the matrix model is
called {\it gaussian}.  A common choice for the Wigner--Dyson
ensembles is

\beq
P_\beta (H)\sim {\rm e}^{-\frac{\beta N}{2v^2}{\rm tr} H^2}
\eeq

where $v$ has the same dimension as the matrix elements of $H$.  The
factor in front of the potential is chosen proportional to $\beta $
for later convenience and proportional to $N$ so that the spectrum,
that has support on some finite interval on the real axis, will remain
bounded in the large $N$ limit.  In this case the macroscopic
eigenvalue spectrum is a semicircle.  Such a spectrum is quite
unrealistic for most physical systems, and not interesting in
itself. What's more, the macroscopic form of the spectrum depends on
the form of the potential. However, the statistical {\it eigenvalue
fluctuations} will turn out to be universal and independent of the
matrix potential in the large--$N$ limit, if we scale the eigenvalues
appropriately (a process referred to as {\it unfolding}). Our focus
will be on such universal quantities.

The probability $P(H)dH$ is invariant under the automorphism

\beq 
\label{eq:rot} 
H\to U^{-1}HU 
\eeq

of the ensemble to itself, where $U$ is a unitary $N\times N$
matrix. By doing an appropriate similarity transformation (\ref{eq:rot}) on
the ensemble of random matrices, the Haar measure $dH$ and potential
$V(H)$ can be expressed in terms of the eigenvectors and eigenvalues
of the matrix $H$:

\beq
\label{eq:ei} 
H\to U^{-1}\Lambda U 
\eeq

where $\Lambda $ is (block)diagonal and contains the eigenvalues. The Haar 
measure is then factorized as follows

\beq
dH=dU\, J(\{ \lambda_i\} )\prod_{i=1}^N d\lambda_i
\eeq

where $dU$ depends only on the eigenvectors and can be integrated out
to give a trivial constant in front of the integral.  $J(\{ \lambda_i\})$ 
is the Jacobian of the similarity transformation (\ref{eq:rot}). It
is given by \cite{Mehta}

\beq
\label{eq:Jbeta}
J_\beta (\{ \lambda_i\} ) \sim \prod_{i<j} |\lambda_i-\lambda_j|^\beta
\eeq

for the Wigner--Dyson ensembles labelled by $\beta $. 

For a gaussian potential $V(H)=\frac{1}{2}H^2$ the partition function 
then takes the form

\eq 
Z_\beta(\{\lambda_i\}) \sim \int P_\beta (\{\lambda_i\})\, 
d\lambda_1...d\lambda_N = 
\int \prod_{i<j} |\lambda_i-\lambda_j|^\beta \,
\e^{-\frac{\beta N}{2v^2}\sum_{j=1}^{N}\lambda_j^2}\, d\lambda_1...d\lambda_N 
\en

This model is easily solvable. This is the
strength of the random matrix description of disordered systems.

The Jacobian in (\ref{eq:Jbeta}) has the form of a Vandermonde
determinant.  If we rewrite eq.~(\ref{eq:Jbeta}) so that the Jacobian
is in the exponent, {\it it gives rise to repulsive eigenvalue
correlations} in the form of a logarithmic pair potential.  Such
correlations are called geometrical, because they arise only from the
Jacobian. In the absence of the invariance (\ref{eq:rot}), the
eigenvalues are uncorrelated and follow a Poisson distribution.

\section{The correlators}
\label{sec-corr}
\setcounter{equation}{0}

The basis for comparing random matrix predictions to experimental or
numerical measurements are the eigenvalue correlation functions. These
determine the statistical properties of the ensemble.

The $k$--point correlation function $\rho_k(\lambda_1,...,\lambda_k)$
is defined as

\beq
\rho_k(\lambda_1,...,\lambda_k)=\frac{N!}{(N-k)!}
\int \prod_{j=k+1}^N d\lambda_jP(\{\lambda_1,...,\lambda_N\})
\eeq

$\rho_k(\lambda_1,...,\lambda_k)d\lambda_1...d\lambda_k$ denotes the
probability of finding any $k$ eigenvalues in the intervals
$d\lambda_i$ around the points $\lambda_i$ ($i=1,...,k$). The 1--point
function is just the density of eigenvalues.

Calculation of the $k$--point correlators can be performed exactly
\cite{Mehta} by rewriting the Jacobian as a product of Vandermonde
determinants of a set of polynomials orthogonal with respect to some
measure $f(\lambda )$. This can easily be done just using the properties of
determinants. For instance, for a gaussian matrix model, $f(\lambda
)={\rm e}^{-\frac{\beta N}{2v^2}\lambda^2}$ and the polynomials are
the Hermite polynomials:

\beq
\int_{-\infty}^{+\infty} H_m(\lambda )H_n(\lambda )\e^{-\frac{\beta N}{2v^2}
\lambda^2}d\lambda=h_n\delta_{mn}
\eeq

where $h_n$ is a normalization. All the classical orthogonal
polynomials appear in this context for various choices of the function
$f(\lambda )$.

The procedure of calculating correlators using orthogonal polynomials
was reviewed in Mehta's book \cite{Mehta}, and was a big step forward
for random matrix theory at the time. The formula for the $k$--point
correlator can be summarized by the formula

\beq 
\rho_{\beta ,k}(\lambda_1,...,\lambda_k)={\rm qdet}[Q_{N\beta}] 
\eeq

where ${\rm qdet}$ denotes a quaternion determinant. \footnote{A
quaternion determinant of an $N\times N$ self--dual quaternion matrix
$Q$ is defined as the square root of the determinant of the $2N\times
2N$ matrix obtained by writing each quaternion as a $2\times 2$
matrix.} The matrix elements of $Q_{N\beta}$ are determined by
universal translation invariant kernels \cite{Mehta} in the large $N$
limit and on the unfolded scale.  Therefore, eigenvalue correlators
are universal and determined only by symmetry. To obtain this
universal behavior, one has to {\it unfold} the spectrum. This means
rescaling the eigenvalues in such a way that they become
dimensionless, by removing the dependence of the non--universal
density of eigenvalues $\rho_1(\lambda )$. Let's define dimensionless
variables $x_i$ by

\beq
\label{eq:defxi_i}
x_i=x_i(\lambda_i )=\int_{-\infty}^{\lambda_i} \rho_1 
(\lambda')d\lambda'
\eeq

The unfolded correlators are obtained by requiring that the
differential probabilities should be the same in the old and in the
new variables:

\beq
\rho_k(\lambda_1,...,\lambda_k)d\lambda_1...d\lambda_k=
X_k(x_1,...,x_k)dx_1...dx_k
\eeq

Note that $X_1(x)=1$ by construction.  

\begin{figure}
\centerline{\psfig{figure=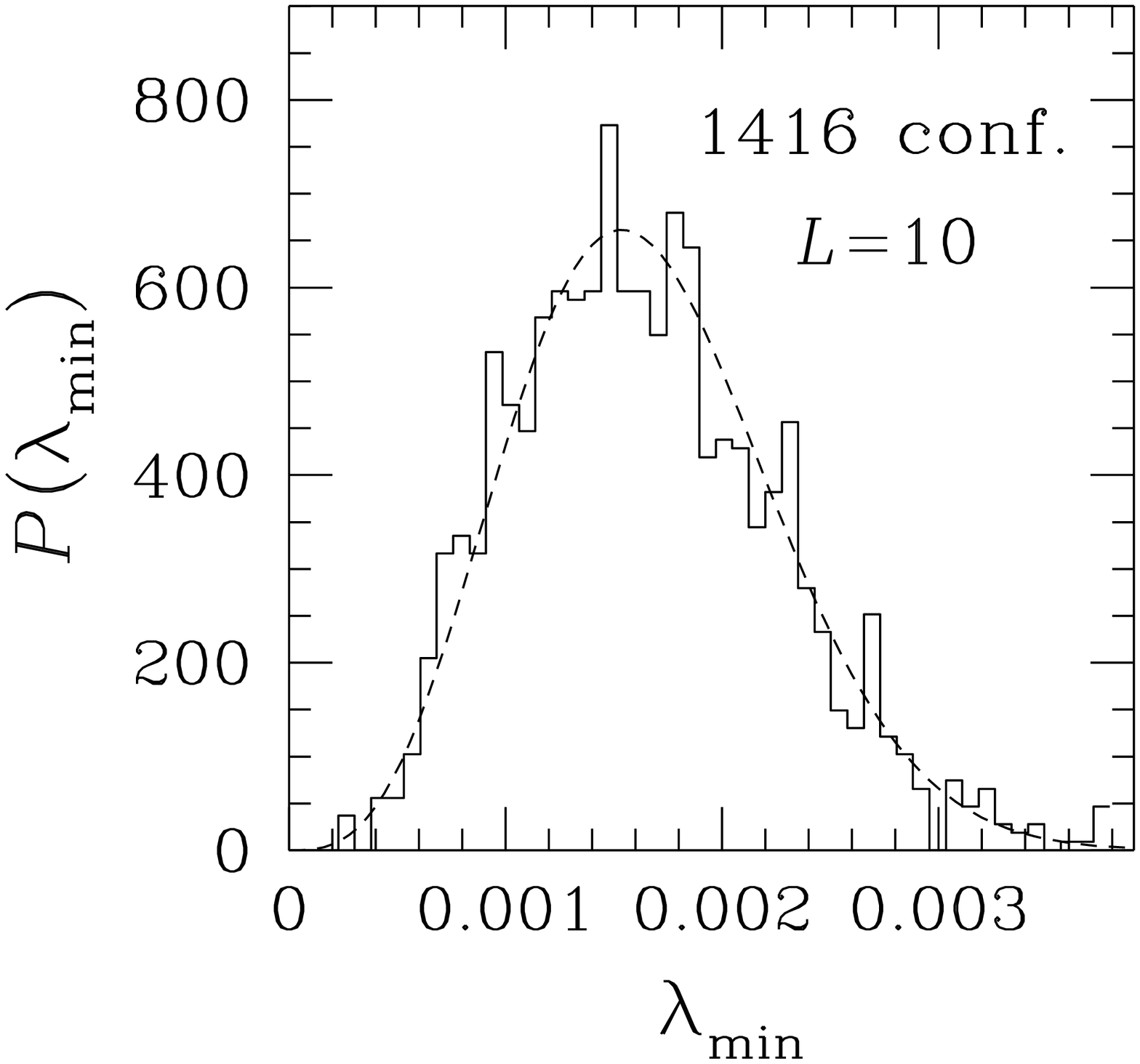,width=50mm}\hspace*{5mm}
  \psfig{figure=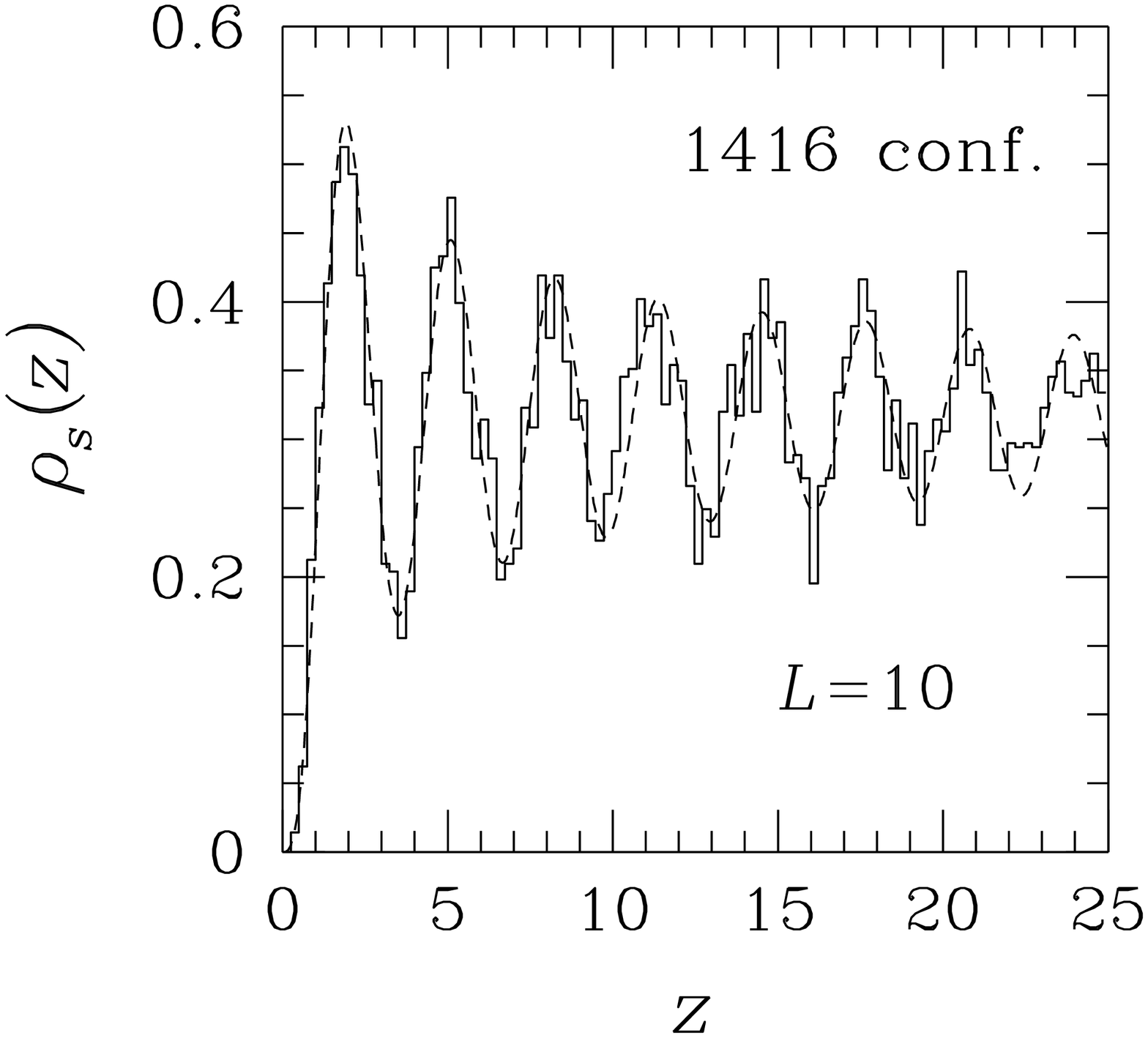,width=50mm}\hspace*{5mm}
  \psfig{figure=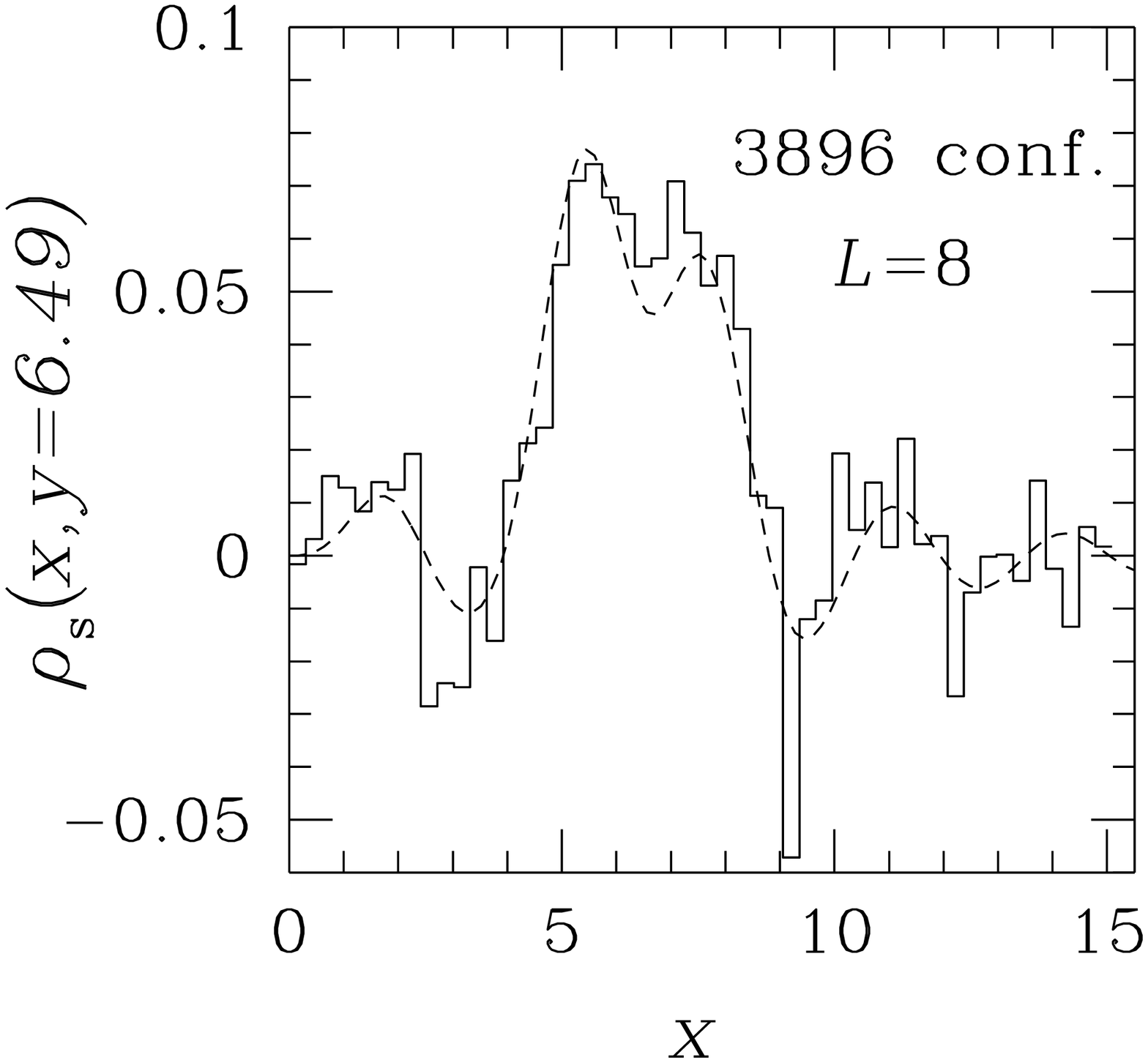,width=50mm}}
\caption{Measurements of the spectrum of the Dirac operator in
SU(2) lattice gauge theory are compared to predictions from random
matrix theory. The figure shows the distribution of the smallest Dirac
eigenvalue $P(\lambda_{\rm min})$, microscopic spectral density
$\rho_S(z)$ (for a 10$^4$ lattice) and the microscopic spectral
two-point function $\rho_S(x,y)$ (for an 8$^4$ lattice).  The
histograms represent lattice SU(2) data; the dashed lines are
analytical predictions from RMT. Reprinted with permission from
ref.~\cite{b-b}. Copyright 2005 by the American Physical Society.}
\label{figlat}
\end{figure}

In practice, however, it is sufficient to do the rescaling in a small
region of the spectrum, where we are interested in studying the
correlations.  If this region is centered, say, at the origin, we put
$z_i=\lambda_i/\Delta$ where $\Delta=\rho_1(0)^{-1}$ is the average
level spacing at the origin (note that this quantity is a function of
$N$). This amounts to magnifying the region on the scale of the
average level spacing, simultaneously with taking the large $N$ limit
in which the spectrum becomes dense. This so--called {\it microscopic}
$k$--point function (we will call it $\rho_{S,k}(z_1,...,z_k)$ in
accordance with standard usage) is then given by

\beq
\rho_{S,k}(z_1,...,z_k)= \begin{array}{c}{\rm lim}\\
{\scriptstyle N\to\infty}\end{array}
\Delta^k \rho_k(\Delta z_1,...,\Delta z_k)
\eeq

where we simultaneously take the limit $N\to \infty$, keeping the
$z_i$ fixed. The new correlators $\rho_{S,k}(z_1,...,z_k)$ are independent of
the random matrix potential within a large class of potentials.
The microscopic spectral correlators can be measured in computer simulations
(see Fig.~\ref{figlat} for an example).

The {\it cluster functions} are relevant for computing spectral
observables. A cluster function is defined as the connected part of a
general $k$--point function. (For comparison, for an uncorrelated
(Poisson) distribution, the connected part vanishes.)  For instance,
before unfolding the two--point function takes the form

\beq
\rho_{2,\beta}(\lambda_1,\lambda_2)=\rho_{1,\beta}(\lambda_1)
\rho_{1,\beta}(\lambda_2)-T_{2,\beta}(\lambda_1,\lambda_2)
\eeq

where the 2--level cluster function is given by
$T_{2,\beta}(\lambda_1,\lambda_2)$.  After unfolding, both the
2--point function and the 2--point cluster function depend only on the
difference $r=x_1-x_2$, and we have

\beq
X_{2,\beta}(r)=1-Y_{2,\beta}(r)
\eeq 

where $X_{2,\beta }(r)$ is the unfolded 2--point function and 
$Y_{2,\beta}(r)$ the unfolded cluster function.

\section{Spectral observables}
\label{sec-stat}
\setcounter{equation}{0}

In comparing random matrix theory with experimental and numerical
results, we need to transform the $k$--point functions into
statistical spectral observables that can be compared to data.

In this paragraph we closely follow ref. \cite{GMW}.  We will consider
an energy spectrum that is the result of a measurement or of a
numerical calculation. The measured values (and the random matrix
eigenvalues) will be denoted by an ordered sequence
$\{E_1,...,E_N\}$. We define a spectral function

\beq
S(E)=\sum_{n=1}^N \delta(E-E_n)
\eeq

and a cumulative spectral function or staircase function

\beq
\eta (E)=\int S(E')dE'=\sum_{n=1}^N \theta(E-E_n)
\eeq

where $\theta$ is the step function. The cumulative spectral function
contains a smooth and a fluctuating part:

\beq
\eta (E)=\xi(E) + \eta_{fl}(E)
\eeq

where $\xi (E)$ is the cumulative mean level density 

\beq
\label{eq:defxi}
\xi=\xi(E)=\int_{-\infty}^E \rho_1(E')dE'
\eeq

\begin{figure}
\centerline{
\psfig{file=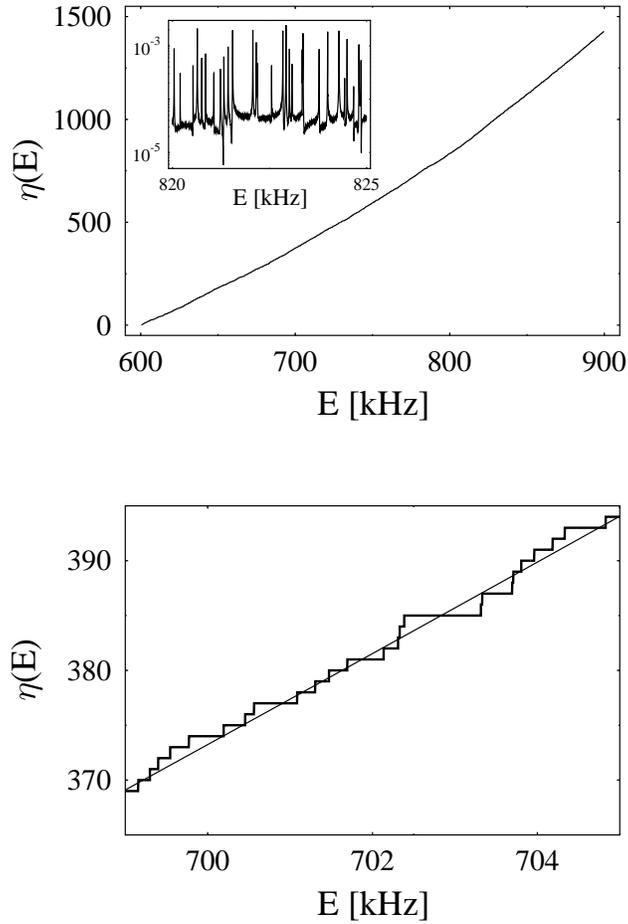,width=3.5in,angle=0}
}
\caption{Example of an experimentally obtained staircase
  function for a spectrum of 1428 elastomechanical eigenfrequencies of
  a resonating quartz block. Due to the high number of levels, the
  staircase function appears as a smooth line. The smooth part
  $\xi(E)$ is a polynomial whose coefficients were found by a fit.
  The bottom part shows a small section of the staircase function.
  Adapted from Ellegaard et al. \cite{elle} by Guhr, M\"uller--Groeling and 
  Weidenm\"uller \cite{GMW}. Copyright 2005 by the 
  American Physical Society and with permission from Elsevier.}
\label{figft8}
\end{figure}

To obtain the smooth part of the experimentally obtained staircase function,
one may fit a polynomial to it (see Fig.~\ref{figft8}).
 
The unfolding procedure is identical to the one we performed in RMT in
the previous paragraph. After unfolding, $\xi_i=\xi(E_i)$, the
staircase function is expressed as

\beq
\hat{\eta } (\xi )=\xi + \hat{\eta }_{fl}(\xi)
\eeq

where $\xi $ is the dimensionless variable defined in
(\ref{eq:defxi}).  Note that the mean level density of the unfolded
spectrum (i.e. the derivative of the smooth part of the step function) is
unity, as we have removed the non--universal dependence.

We will now discuss how a few spectral observables are obtained
from the correlation functions.

To study long--range correlations a common statistic is the {\it level
number variance} $\Sigma^2(L)$. If $\hat{\eta}$ denotes
the number of levels in the interval $[\xi,\xi+L]$ in the unfolded
spectrum, the number variance is defined by

\beq
\Sigma^2(L)=\langle \hat{\eta}^2\rangle - 
\langle \hat{\eta} \rangle^2=
\langle \hat{\eta}^2\rangle - L^2
\eeq

where the angular brackets $\langle ... \rangle $ denote the
average with respect to $\xi$. In an interval of length $L$
one expects on average $L\pm \sqrt{\Sigma^2(L)}$ levels.

The number variance is given in terms of correlators by

\beq
\Sigma_\beta^2(L)= L-2\int_0^L (L-r)Y_{2,\beta}(r)\, dr
\eeq

where $Y_{2,\beta}(r)$ is the unfolded 2--level cluster function.  

Another long--range statistic is the {\it spectral rigidity}
$\Delta_3$. It is defined as the least square deviation of the
unfolded cumulative spectral function (staircase function) from the
best fit to a straight line:

\beq
\Delta_3(L)=L^{-1}{\Big \langle} \begin{array}{c}{\rm min} \\ a,b\end{array}
\int_{\xi}^{\xi+L} {\Big (}\hat{\eta}(\xi')-(a\xi'+b){\Big )}^2\, 
d\xi' 
{\Big \rangle}
\eeq

where, like in the number variance, the angular brackets $\langle
... \rangle $ denote the average with respect to $\xi$. The spectral
rigidity can similarly to $\Sigma^2(L)$ be expressed as an integral
involving $Y_2(r)$.

\begin{figure}
\centerline{
\psfig{file=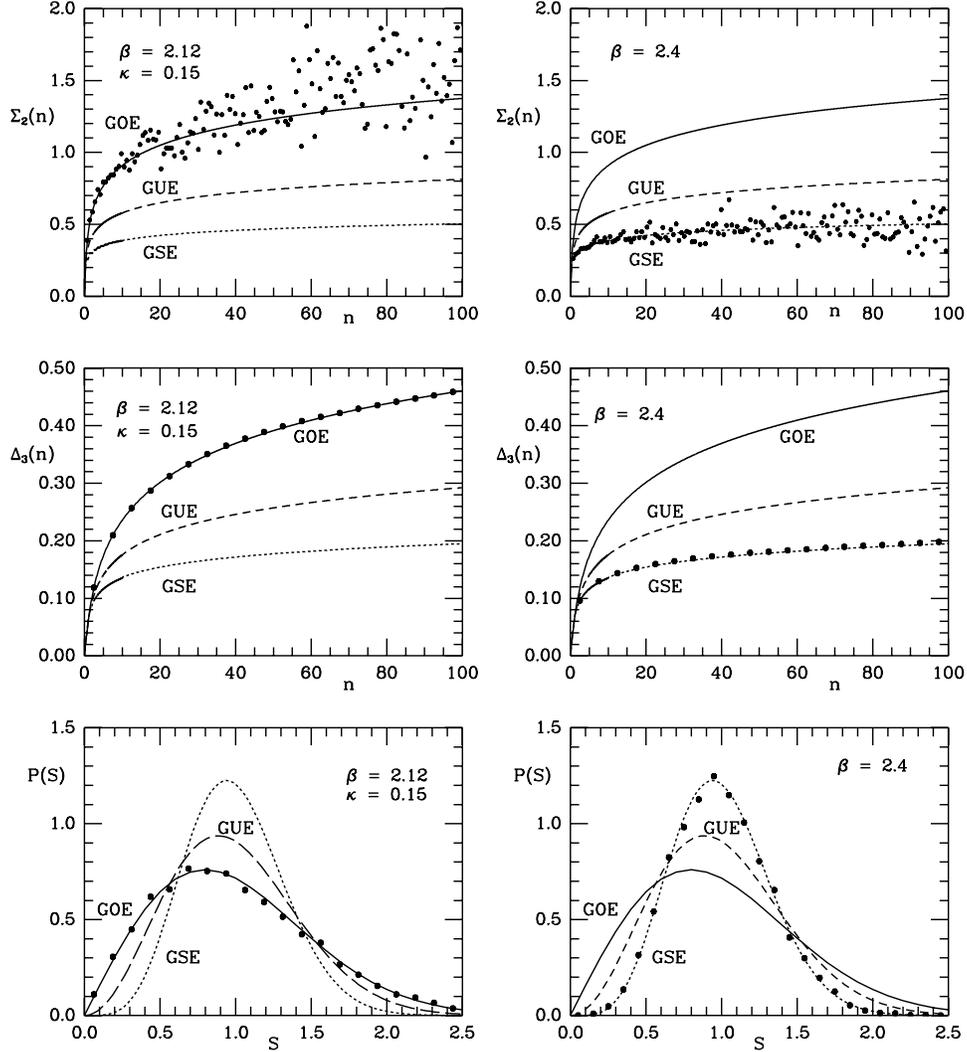,width=5in}}
\caption{Statistical observables in the Dirac spectrum of SU(2) gauge
theory.  The figures show a comparison of lattice data and RMT
predictions for the number variance $\Sigma_2$, the $\Delta_3$
statistic, and the nearest neighbor spacing distribution $p(s)$. The
two columns show different implementations of fermions on the lattice
(left: Wilson fermions, right: Kogut-Susskind fermions). GOE, GUE and
GSE stand for the Gaussian Orthogonal, Unitary, and Symplectic
Ensemble, respectively. Reprinted from \cite{Jac_P3} with permission
from Elsevier.}
\label{fig9}
\end{figure}

To study fluctuations in the spectrum on a short scale (a few level
spacings) we can study the {\it nearest neighbor spacing distribution}
$p(s)$.  This is the probability density for two neighboring levels
$\xi_n$ and $\xi_{n+1}$ being a distance $s$ apart. The calculation of
$p(s)$ is non--trivial and involves all correlation functions $\rho_k$
with $k\geq 2$. An excellent approximation is given by the Wigner
surmise: 

\beq
p_\beta(s)=a_\beta s^\beta {\rm exp}(-b_\beta s^2)
\eeq

where $a_\beta $, $b_\beta $ are $\beta $--dependent constants. Note
the level repulsion factor $s^\beta $ at small $s$. See Fig.~\ref{fig9}
for an example of measurements of these statistical observables.

For comparison of measurements to RMT predictions to make sense, we
have to make the assumption of ergodicity. This means that we assume
that the ensemble average in the theoretical prediction of RMT is
equal to the running average over the sequence of measurements on a
single sample:

\beq
\langle f(E) \rangle_{ens}=\langle f(E) \rangle_{meas} 
\eeq

where $f(E)$ denotes any function of the eigenvalues. 

The observed spectral fluctuations in the systems we have discussed 
in the introduction show, in many instances, an impressive agreement
with random matrix theory predictions.

\newpage
\section{Lie groups, algebras, and root lattices}
\label{sec-lie}
\setcounter{equation}{0}

In this lecture we will first present some preliminary material leading
up to the definition of symmetric spaces. Assuming that most of
the audience is more familiar with this concept than the average
physicist, we will be as brief as possible.

As already mentioned, the reason we are interested in symmetric spaces
in connection with RMT is that random matrix ensembles are identified
with symmetric spaces. As we will see, symmetric spaces (SS) have
well--known properties \cite{Helgason,Helgason2} and much of the
theory for SS can be used in physical problems where RMT is applicable.
We will give a few examples of such usage in the next lecture.

We will start by reminding the reader of some basic definitions
concerning Lie algebras and root spaces.
A Lie algebra ${\bf G}$ is a linear vector space over a field $F$. 
Multiplication in the Lie algebra is given by the Lie bracket
$[X,Y]$. It has the following properties:

\noindent [1] If $X$, $Y\in {\bf G}$, then $[X,Y]\in {\bf G}$,\\
\noindent [2] $[X,\alpha Y+\beta Z]=\alpha [X,Y]+\beta[X,Z]$ for $\alpha $,
$\beta \in F$, \\
\noindent [3] $[X,Y]=-[Y,X]$, \\
\noindent [4] $[X,[Y,Z]]+[Y,[Z,X]]+[Z,[X,Y]]=0$ (the Jacobi identity).

The algebra ${\bf G}$ generates a group through the exponential mapping.
A general group element is 

\beq
M={\rm exp}\left( \sum_it^iX_i\right);\ \ \ \ t^i\in F,\ X_i\in {\bf G}
\eeq

where $t_i$ are parameters (coordinates).  We define a mapping ${\rm
ad} X$ from the Lie algebra to itself by ${\rm ad} X:Y\to [X,Y]$.  The
mapping $X\to {\rm ad} X$ is a representation of the Lie algebra
called the adjoint representation.  It is easy to check that it is an
automorphism (i.e. that it preserves algebraic operations): it follows
from the Jacobi identity that $[{\rm ad}X_i,{\rm ad}X_j]={\rm
ad}[X_i,X_j]$.  Suppose we choose a basis $\{ X_i\}$ for ${\bf
G}$. Then

\beq
\label{eq:adjr}
{\rm ad} X_i(X_j)=[X_i,X_j]=C^k_{ij}X_k
\eeq

where we sum over $k$. The $C^k_{ij}$ are real structure
constants. The structure constants define the matrices $M$ of the
adjoint representation through $(M_i)_{jk}=C^j_{ik}$.

An {\it ideal} ${\bf I}$ is a subalgebra such that $[{\bf G},{\bf I}]
\subset {\bf I}$.  A {\it simple} Lie algebra has no proper ideal.
The semisimple algebras are built from the simple ones. In any simple
algebra there are two kinds of generators.

(1) There is a maximal abelian subalgebra, called the {\it Cartan
subalgebra} ${\bf H_0}= \{ H_1,...,H_r\} $ such that

\beq
[H_i,H_j]=0 
\eeq

If we represent each element of the Lie algebra by an $n\times n$
matrix, then $[H_i,H_j]=0$ means the matrices $H_i$ can all be
diagonalized simultaneously. Their eigenvalues $\mu_i$ are given by

\beq
H_i|\mu \rangle =\mu_i|\mu \rangle
\eeq

where the eigenvectors are labelled by the {\it weight vectors} $\mu
=(\mu_1,...,\mu_r)$. A {\it positive weight} is a weight whose first
non--zero component is positive.

(2) There are raising and lowering operators denoted
$E_\alpha$ such that 

\beq
[H_i,E_\alpha]=\alpha_iE_\alpha, \ \ \ \ \ 
[E_\alpha,E_{-\alpha}]=\alpha_iH_i
\eeq

Here $\alpha $ is an $r$--dimensional vector called a {\it root}:
$\alpha =(\alpha_1,...,\alpha_r)$ and $r$ is the {\it rank} of the
algebra. For each root $\alpha_i$, there is another root $-\alpha_i$
and a corresponding eigenoperator $E_{-\alpha}$. The roots form a
lattice in the space dual to the Cartan subalgebra. A subset of the
positive roots span the root lattice. These are the {\it simple
roots}. Their number is equal to $r$, the rank of the algebra.
All the weights of a representation can be obtained by acting on the
highest weight with lowering operators in all possible ways.  

One can prove the following relation between roots and weights:

\beq
\label{eq:fund}
\frac{2\alpha \cdot \mu}{\alpha^2}=-(p-q)
\eeq

where $p$, $q$ are positive integers such that $E_\alpha |\mu +p\alpha
\rangle =0$, $E_{-\alpha} |\mu -q\alpha \rangle =0$, i.e. they define
the distance of $|\mu \rangle$ to the upper and lower end of the
ladder.

Eq.~(\ref{eq:fund}) implies that the possible angle between two root
vectors of a simple Lie algebra is limited to multiples of $\frac{\pi
}{6}$ and $\frac{\pi }{4}$. Therefore, there is a finite set of
possible root lattices. Equation (\ref{eq:fund}) permits a
classification of all complex semisimple algebras. The classical Lie
algebras ${\bf SU(n+1,C)}$, ${\bf SO(2n+1,C)}$, ${\bf Sp(2n,C)}$ and
${\bf SO(2n,C)}$ correspond to root systems $A_n$, $B_n$, $C_n$, and
$D_n$, respectively.  In addition there are five exceptional
algebras, but these are not relevant for random matrix theory because 
they have a finite $n$. 

The root systems for these four infinite series of classical
non--exceptional Lie groups can be characterized as follows
\cite{Georgi} (denote the $r$--dimensional space spanned by the roots
by ${\cal V}$ and let $\{e_1,...e_n\}$ be a canonical basis in ${\bf
R}^n$):

$A_{n-1}$: Let ${\cal V}$ be the hyperplane in ${\bf R}^n$ that passes
through the points $(1,0,0,...0)$, $(0,1,0,...,0)$, ...,
$(0,0,...,0,1)$ (the endpoints of the $e_i$, $i=1,...,n$). Then the
root lattice contains the vectors $\{ e_i-e_j, i\neq j \}$.

$B_n$:  Let ${\cal V}$ be ${\bf R}^n$; then the roots are $\{ \pm e_i,
\pm e_i \pm e_j, i\neq j \}$.

$C_n$: Let ${\cal V}$ be ${\bf R}^n$; then the roots are $\{ \pm 2e_i,
\pm e_i \pm e_j, i\neq j \}$.

$D_n$: Let ${\cal V}$ be ${\bf R}^n$; then the roots are $\{
\pm e_i \pm e_j, i\neq j \}$.

The root lattice $BC_n$, that we will discuss in conjunction with restricted
root systems, is the union of $B_n$ and $C_n$. It is characterized as follows:

$BC_n$:  Let ${\cal V}$ be ${\bf R}^n$; then the roots are $\{ \pm e_i, 
\pm 2e_i, \pm e_i \pm e_j, i\neq j \}$. 

Roots of length 1, $\sqrt{2}$, and 2 are called {\it short, ordinary,
and long roots}, respectively.  Each of the complex algebras in
general has several {\it real forms} associated with it.\footnote{Also
symmetric spaces have real forms, but they will not be discussed
here. These are pseudo--riemannian symmetric spaces, i.e. they have a
non--definite metric.} We will define these
shortly. Eq.~(\ref{eq:fund}) expresses invariance of the root lattice
under reflections in the hyperplanes orthogonal to the roots (the Weyl
group). If $\mu $ is a weight or root, so is $\mu'$:

\beq
\label{eq:Weyl}
\mu'=\mu - \frac{2(\alpha \cdot \mu )}{\alpha^2}\alpha
\eeq

The relation (\ref{eq:fund}) determines the highest weights of
all irreducible representations. Setting $p=0$, choosing a positive
integer $q$, and letting $\alpha $ run through the simple roots,
$\alpha=\alpha^i$ ($i=1,...,r$), we find the highest weights $\mu^i $
of all the irreducible representations corresponding to the given
value of $q$ \cite{Georgi}.

\section{Cosets}
\label{sec-cosets}
\setcounter{equation}{0}

In general, a symmetric space can be represented as a coset space of
some Lie group $G$ with respect to a symmetric subgroup $H$. 
The (left) coset space $G/H$ is the set of subsets of $G$ of the form $gH$
($g\in G$):

\beq
G=g_0H+g_1H+...+g_nH
\eeq

Every element $g\in G$ can be written uniquely as $g=g_ih_j$ for some
$g_i\in G$ and some $h_j\in H$.  The coset can be identified with the
set of group operations $\{g_0,...,g_n\}$.  The coset corresponds to a
manifold of dimension ${\rm dim}G-{\rm dim}H$, as we will see in the
example below.

Suppose $G$ is represented by matrices acting transitively on a space
$V$ ($Gv=V$ for any $v\in V$) and $Hv_0=v_0$ for some $v_0$ (then $H$
is called the isotropy subgroup at the point $v_0$). Then there is
one--to--one correspondence between the elements in $V$ and those in
$G/H$: $gHv_0=gv_0=v$.

{\bf Example:} The $SO(2)$ subgroup of $SO(3)$ is the isotropy
subgroup at the north pole of a unit 2--sphere imbedded in
3--dimensional space, since it keeps this point fixed. On the
other hand, the north pole is mapped onto any point on the surface of
the sphere by elements of the coset $SO(3)/SO(2)$.

The ${\bf SU(3)}$ algebra is defined by the commutation relations

\beq
[L_i,L_j]=\frac{1}{2}\epsilon_{ijk}L_k 
\eeq

where $\frac{1}{2}\epsilon_{ijk}$ are structure constants. A matrix 
representation of this algebra is given by 

\beq 
L_1= \frac{1}{2}\left(\begin{array}{ccc} 0&0&0\\ 0&0&1\\
0&-1&0\end{array}\right), \ \ \ \
L_2=\frac{1}{2}\left(\begin{array}{ccc} 0&0&1\\ 0&0&0\\
-1&0&0\end{array}\right), \ \ \ \
L_3=\frac{1}{2}\left(\begin{array}{ccc} 0&1&0\\ -1&0&0\\
0&0&0\end{array}\right)
\eeq

The subgroup $SO(2)$ is generated by $L_3$. This subgroup keeps the
north pole fixed:

\beq
\label{eq:fixedNP}
{\rm exp}(t^3L_3)\left(\begin{array}{c} 0 \\ 0 \\ 1 \end{array}\right)=
\left(\begin{array}{c} 0 \\ 0 \\ 1 \end{array}\right)
\eeq

The remaining group generators define the coset space $SO(3)/SO(2)$. In terms of the 
real coordinates $t^1$, $t^2$, an element in this coset space takes the form

\beq
\begin{array}{l}
\label{eq:L1L2}
M={\rm exp}\left(\sum_{i=1}^2 t^iL_i\right) \nonumber \\ \nonumber \\
=\left(
  \begin{array}{ccc} 1+(t^2)^2\frac{({\rm cos}\sqrt{(t^1)^2+(t^2)^2}-1)}{(t^1)^2+(t^2)^2} &
                     t^1t^2\frac{({\rm cos}\sqrt{(t^1)^2+(t^2)^2}-1)}{(t^1)^2+(t^2)^2} &
                     t^2\frac{{\rm sin}\sqrt{(t^1)^2+(t^2)^2}}{\sqrt{(t^1)^2+(t^2)^2}}\\

                     t^1t^2\frac{({\rm cos}\sqrt{(t^1)^2+(t^2)^2}-1)}{(t^1)^2+(t^2)^2} &
                     1+(t^1)^2\frac{({\rm cos}\sqrt{(t^1)^2+(t^2)^2}-1)}{(t^1)^2+(t^2)^2} &
                     t^1\frac{{\rm sin}\sqrt{(t^1)^2+(t^2)^2}}{\sqrt{(t^1)^2+(t^2)^2}}\\

                     -t^2\frac{{\rm sin}\sqrt{(t^1)^2+(t^2)^2}}{\sqrt{(t^1)^2+(t^2)^2}} &
                     -t^1\frac{{\rm sin}\sqrt{(t^1)^2+(t^2)^2}}{\sqrt{(t^1)^2+(t^2)^2}} &
                     {\rm cos}\sqrt{(t^1)^2+(t^2)^2}\end{array}
\right)\nonumber \\ \nonumber \\
\equiv \left(\begin{array}{ccc}  .   &  .   & x \\
                            .   &  .   & y \\
                            .   &  .   & z \end{array}\right);\ \ \ \ \ x^2+y^2+z^2=1
\end{array} \nonumber 
\eeq

The last equation is the equation for a 2--sphere.  When the coset
space representative $M$ acts on the north pole the orbit is exactly
the 2--sphere:

\beq
M\left(\begin{array}{c} 0\\ 0\\ 1\end{array}\right) =
\left(\begin{array}{ccc}  .   &  .   & x \\
                          .   &  .   & y \\
                          .   &  .   & z \end{array}\right)
\left(\begin{array}{c} 0\\ 0\\ 1\end{array}\right) =
\left(\begin{array}{c} x\\ y\\ z\end{array}\right) 
\eeq

Because of this one--to--one correspondence, the coset space
$SO(3)/SO(2)$ can be identified with a unit 2--sphere imbedded in
3--dimensional space. 

\section{Symmetric spaces}
\label{sec-ss}
\setcounter{equation}{0}

Suppose ${\bf G}$ is a compact simple Lie algebra.  A linear
automorphism $\sigma \neq 1$ of the Lie algebra ${\bf G}$ onto itself
such that $\sigma^2=1$ is called an {\it involutive automorphism} or
{\it involution}. This means the eigenvalues of $\sigma $ are $\pm 1$,
and $\sigma $ splits the algebra ${\bf G}$ into orthogonal
eigensubspaces corresponding to these eigenvalues: ${\bf G}={\bf K}
\oplus {\bf P}$ where

\beq
\label{eq:KP}
\sigma (X)=X\ \  {\rm for}\ \  X\in {\bf K},\ \ \sigma (X)=-X\ \  
{\rm for}\ \  X\in {\bf P}
\eeq

${\bf K}$ is a subalgebra, but ${\bf P}$ is not. From
eq.~(\ref{eq:KP}), the following commutation relations hold:

\beq
\label{eq:commrel}
[{\bf K},{\bf K}]\subset {\bf K},\ \  [{\bf K},{\bf P}]\subset {\bf P},\ \  
[{\bf P},{\bf P}]\subset {\bf K} 
\eeq

A subalgebra ${\bf K}$ satisfying (\ref{eq:commrel}) is called 
{\it symmetric}.  If we now multiply the elements in ${\bf
  P}$ by $i$ (this is called the ``Weyl unitary trick''), we construct
a new noncompact algebra ${\bf G^*}={\bf K} \oplus i{\bf P}$. This is
called a {\it Cartan decomposition} of ${\bf G^*}$, and ${\bf K}$ is a
maximal compact subalgebra.  The coset spaces $G/K$ and $G^*/K$ are
{\it symmetric spaces} of compact and non--compact type, respectively.

{\bf Example}: Suppose $G=SU(3,C)$, the group of $3\times 3$ unitary
complex matrices with unit determinant. The algebra of this group
consists of eight complex antihermitean traceless matrices $X_i$,
$i=1,...,8$. Let us take a representation $X_i=iT_i$ where $T_i$
denote the Gell--Mann matrices known to physicists.

An involution that splits the ${\bf SU(3,C)}$ algebra in two subspaces
${\bf K}$, ${\bf P}$ defined as above is given by complex conjugation
$\sigma=K$.  This involution splits the algebra $\{X_1,...X_8\}$ into
real and pure imaginary matrices. In the Gell--Mann representation,

\beq
\label{eq:SU3}
\begin{array}{l}
{\bf K}=\{ X_2,X_5,X_7\}
=\left\{
\frac{1}{2}\left(\begin{array}{ccc} 0&1&0\\ -1&0&0\\ 0&0&0\end{array}\right),
\frac{1}{2}\left(\begin{array}{ccc} 0&0&1\\ 0&0&0\\ -1&0&0\end{array}\right),
\frac{1}{2}\left(\begin{array}{ccc} 0&0&0\\ 0&0&1\\ 0&-1&0\end{array}\right)
\right\}\\
\\
{\bf P}=\{ X_1,X_3,X_4,X_6,X_8\}\\ \\
=\left\{
\frac{i}{2}\left(\begin{array}{ccc} 0&1&0\\1&0&0\\0&0&0\end{array}\right),
\frac{i}{2}\left(\begin{array}{ccc} 1&0&0\\0&-1&0\\0&0&0\end{array}\right),
\frac{i}{2}\left(\begin{array}{ccc} 0&0&1\\0&0&0\\1&0&0\end{array}\right),
\frac{i}{2}\left(\begin{array}{ccc} 0&0&0\\0&0&1\\0&1&0\end{array}\right),
\frac{i}{2\sqrt{3}}\left(\begin{array}{ccc} 1&0&0\\0&1&0\\0&0&-2\end{array}\right)
\right\} \end{array} \nonumber \\ 
\nonumber \\ 
\eeq

${\bf K}$ is the compact subalgebra ${\bf SO(3,R)}$
consisting of real, skew--symmetric and traceless matrices
(this can easily be checked by putting $X_2 \equiv L_3$, $X_5 \equiv
L_2$, $X_7 \equiv L_1$ and comparing with the $SO(3)$ commutation relations
$[L_i,L_j]=\frac{1}{2}\epsilon_{ijk}L_k$), and ${\bf
  P}$ is the subspace of matrices of the form $iT$, where $T$ is real,
symmetric, and traceless.  The Cartan subalgebra is given by
$\{X_3,X_8\}\subset {\bf P}$.

By the Weyl unitary trick we now obtain from ${\bf G}$ the
non--compact algebra ${\bf G^*}={\bf K}\oplus i{\bf P}$, where $i{\bf
P}$ is a subspace of real, symmetric, and traceless matrices $-T$. The
entire Lie algebra ${\bf G^*}$ consists of $3\times 3$ real matrices
of zero trace, and generates the linear group of transformations
$SL(3,R)$.

The coset space $G/K=SU(3,C)/SO(3,R)$ is a symmetric space of compact
type, and the related symmetric space of non--compact type is
$G^*/K=SL(3,R)/SO(3,R)$. 

Note that the tangent space of $G/K$ (or $G^*/K$) at the origin
(identity element) is spanned by the subspace ${\bf P}$ (or $i{\bf
P}$, respectively) of the algebra.  Let's denote by $P={\rm e}^{\bf
P}$ the exponential of any point in the algebra subspace spanned by
the set ${\bf P}$ (such a point is a linear combination of the
generators in this subspace).  When $K$ is a connected subgroup, $P$
is isomorphic to $G/K$. In general $P$ is not a subgroup. However, one
can show that if $p\in P$, then also $p'=kpk^{-1}\in P$. This defines
a transitive group action on $P$. Also, if $K$ is compact, every $p\in
P$ is conjugate with some element in the Cartan subalgebra:

\beq
\label{eq:sphdec}
p=khk^{-1}
\eeq

This is called spherical decomposition.  It defines the {\it
angular} coordinate $k$ and the {\it spherical radial} coordinate $h$
of the point $p\in P$. In plain language, every matrix in the coset
space $G/K$ or $G^*/K$ can be diagonalized by a similarity
transformation by the subgroup $K$.

{\bf Example:} In the adjoint representation, which has the same
dimension as the group, the complex symmetric matrices in $G/K$
$=SU(3,C)/SO(3)$ $\simeq P=\e^{\bf P}$ can be diagonalized by the
group $K=SO(3)$ to the form

\beq
\label{eq:adjrad}
p=khk^{-1}; \ \ \ \ h=\e^{i{\bf t\cdot H}}=
=\left(\begin{array}{ccccccc} 1 & ...    &  & & & & \\               
                   . & 1 &   & & & &\\
                   . &   & \e^{i{\bf t\cdot \alpha }} & & & &\\
                     &   &   & \e^{-i{\bf t\cdot \alpha }} & & & \\
                     &   &   &   & \e^{i{\bf t\cdot \beta }} & & \\

                     &   &   &   &    & \ddots & \\
                     &        &   &   &       &        & 
\e^{-i{\bf t\cdot \gamma }} \end{array}\right)
\eeq

where we have written the factor of $i$ multiplying the generators in
the Cartan subalgebra explicitly in the exponent ($X_3\equiv iH_1, \
X_8\equiv iH_2$).  The vectors $\pm {\bf \alpha },\ \pm {\bf \beta },\
\pm {\bf \gamma }$ are the three pairs of roots of $SU(3)$ (these form
a regular hexagon in the plane) and the diagonal elements equal to 1
are the exponentials of the zero roots corresponding to the two
operators in the Cartan subalgebra.

\section{The metric on a Lie algebra}  
\label{sec-metric}
\setcounter{equation}{0}

A metric tensor can be defined on a Lie algebra.  This will be useful
for defining the curvature of symmetric spaces.  Let $\{ X_i\}$ be a
basis for the Lie algebra ${\bf G}$ and let $C_{ij}^k$ denote the
structure constants in this basis. The metric tensor on the algebra may
be defined by the {\it Killing form} $K(X_i,X_j)$

\beq
\label{eq:metric}
g_{ij}=K(X_i,X_j)\equiv \tr ({\rm ad}X_i {\rm ad}X_j)=C^r_{is}C^s_{jr}
\eeq

The Killing form is symmetric and bilinear. According to Cartan, the
Killing form is non--degenerate for a semisimple algebra.  This means
that ${\rm det}g_{ij}\neq 0$, so that the inverse of $g_{ij}$, denoted
by $g^{ij}$, exists. Since it is also real and symmetric, it can be
reduced to canonical diagonal form $g_{ij}={\rm diag}(-1,...,-1,1,...,1)$.

According to a theorem by Weyl, a simple Lie group $G$ is
compact, if and only if the Killing form on ${\bf G}$ is negative
definite. Otherwise it is non--compact.

{\bf Example:} We have already written down the commutation relations
of the compact group $SO(3,R)$ in the form
$[L_i,L_j]=C_{ij}^kL_k=\frac{1}{2}\epsilon_{ijk}L_k$. We can
renormalize the generators so that the entries of the metric are
unity. The commutation relations then take the form

\beq
\label{eq:tildeSigmarels}
\begin{array}{c} 
[\tilde{L}_1,\tilde{L}_2]=-\frac{1}{\sqrt 2}\tilde{L}_3,\ \ \ \ \ \ \ 
[\tilde{L}_2,\tilde{L}_3]=-\frac{1}{\sqrt 2}\tilde{L}_1,\ \ \ \ \ \ \ 
[\tilde{L}_3,\tilde{L}_1]=-\frac{1}{\sqrt 2}\tilde{L}_2\end{array}
\eeq

We can read off the structure constants and then, using
eq.~(\ref{eq:metric}), compute the components of the Killing form. In
this normalization it is

\beq
g_{ij}=\left(\begin{array}{ccc} -1 &    &    \\
                                   & -1 &    \\
                                   &    & -1 \end{array}\right)
\eeq

(In the unrenormalized form it is $g_{ij}=-\frac{1}{2}\delta_{ij}$.)
This is the metric in the $SO(3)$ {\it algebra}. It is negative,
because the group is compact.

The generators of the non--compact group $SO(2,1;R)$ obey the
commutation relations

\beq
\label{eq:Sigmarels}
\begin{array}{c} 
[\Sigma_1,\Sigma_2]=-\frac{1}{\sqrt 2}\Sigma_3,\ \ \ \ \ \ \ 
[\Sigma_2,\Sigma_3]=-\frac{1}{\sqrt 2}\Sigma_1,\ \ \ \ \ \ \ 
[\Sigma_3,\Sigma_1]=\frac{1}{\sqrt 2}\Sigma_2\end{array}
\eeq

In the same way, using again eq.~(\ref{eq:metric}) we compute the
matrix elements $g_{ij}$. The result is

\beq
g_{ij}=\left(\begin{array}{ccc} 1 &    &    \\
                                   & 1 &    \\
                                   &    & -1 \end{array}\right)
\eeq

we have labelled the rows and columns in the order 3,1,2. The 
generator $\Sigma_2$ makes up the compact subalgebra ${\bf K}$.

\section{The metric on a symmetric space}  
\label{sec-metricSS}
\setcounter{equation}{0}

The definition of the metric can be extended to an arbitrary point of
a symmetric space.  At the origin, the metric is defined by
restricting the metric in the algebra to the tangent space (remember
that the latter is spanned by the generators in the subspace ${\bf P}$
or $i{\bf P}$ of the whole algebra ${\bf G}$ or ${\bf G^*}$). 

{\bf Example:} The metric in the subspace ${\bf P}$ of ${\bf SO(3,R)}$
is obtained by excluding the row and column corresponding to the generator
in ${\bf K}$, keeping the ones in ${\bf P}$:

\beq
g_{ij}=\left(\begin{array}{cc} -1 &      \\
                                  & -1    \end{array}\right)
\eeq

Similarly, the metric in the subspace $i{\bf P}$ of ${\bf SO(2,1;R)}$
is obtained by excluding the row and column corresponding to the 
compact generator in ${\bf K}$ and keeping the ones in $i{\bf P}$:

\beq
g_{ij}=\left(\begin{array}{cc} 1 &     \\
                                 & 1   \end{array}\right)
\eeq

Since the group acts transitively on the symmetric space, we can then
use a group transformation to map the metric to an arbitrary point of
the SS, using the invariance of the line element in local coordinates
given by $ds^2=g_{ij}dx^idx^j$. 

Note that if a positive metric is required on a compact symmetric
space, we can use minus the Killing form, which sometimes is more
natural.

{\bf Example:} The line element $ds^2$ on the radius--1 2--sphere
isomorphic to the symmetric space $SO(3,R)/SO(2)$ in polar coordinates
is $ds^2= d\theta^2+{\rm sin}^2\theta \, d\phi^2$. The metric at the
point $(\theta,\phi)$ is

\beq
\label{eq:metric_on_sphere}
g_{ij}=\left(\begin{array}{cc} 1 & 0 \\
                       0 & {\rm sin}^2\theta \end{array}\right),\ \ \ \ \ \ \  
g^{ij}=\left(\begin{array}{cc} 1 & 0 \\
                       0 & {\rm sin}^{-2}\theta \end{array}\right) 
\eeq

where the rows and columns are labelled in the order $\theta $, $\phi $.

The line element $ds^2$ on the hyperboloid $SO(2,1;R)/SO(2)$ in polar
coordinates is $ds^2= d\theta^2+{\rm sinh}^2\theta \, d\phi^2$. The
metric at the point $(\theta,\phi)$ is

\beq
\label{eq:metric_on_hyp}
g_{ij}=\left(\begin{array}{cc} 1 & 0 \\
                       0 & {\rm sinh}^2\theta \end{array}\right),\ \ \ \ \ \ \  
g^{ij}=\left(\begin{array}{cc} 1 & 0 \\
                       0 & {\rm sinh}^{-2}\theta \end{array}\right) 
\eeq

\section{Real forms and the metric}
\label{sec-realmet}
\setcounter{equation}{0}

The form of the metric depends on the basis of the algebra.
A complex Lie algebra ${\bf G^C}$ is given by

\beq
\label{eq:complexalg}
{\bf G^C}=
\sum_i c^iH_i + \sum_\alpha c^\alpha E_\alpha \ \ \ \ \ \ \ (c^i,\ c^\alpha 
\ {\rm complex})
\eeq

where ${\bf H_0}=\{ H_i\} $ is the Cartan subalgebra and $\{E_{\pm
\alpha }\}$ are the pairs of raising and lowering operators.  A {\it
real form} of the same algebra is obtained by taking the coordinates
$c^i$, $c^\alpha$ to be real numbers, i.e.

\beq
\label{eq:realalg}
{\bf G}=
\sum_i c^iH_i + \sum_\alpha c^\alpha E_\alpha \ \ \ \ \ \ \ (c^i,\ c^\alpha 
\ {\rm real})
\eeq

We can choose different basis vectors in this real algebra. The form 
of the metric will change accordingly.

The metric corresponding to the basis $\{H_i,\pm E_\alpha\}$ is not diagonal.
It has the form

\beq
\label{eq:raislow_metric}
g_{ij}=\left( \begin{array}{cccccccc}  1 &        &    &   &    &        &   &    \\
                                         & \ddots &    &   &    &        &   &    \\
                                         &        &  1 &   &    &        &   &    \\
                                         &        &    & 0 & 1  &        &   &    \\
                                         &        &    & 1 & 0  &        &   &    \\
                                         &        &    &   &    & \ddots &   &    \\
                                         &        &    &   &    &        & 0 & 1  \\
                                         &        &    &   &    &        & 1 & 0  \end{array} \right)
\eeq

By recombining the basis vectors into

\beq
\label{eq:NRFd}
{\bf K}=\left\{\frac{(E_\alpha - E_{-\alpha})}{\sqrt{2}}\right\},\ \ \ \ \ \ \ 
i{\bf P}=\left\{H_i,\frac{(E_\alpha + E_{-\alpha})}{\sqrt{2}}\right\} 
\eeq

the metric takes the diagonal form

\beq
\label{eq:NRFd_metric}
g_{ij}=\left( \begin{array}{cccccccc}  1 &        &    &   &    &        &   &    \\
                                         & \ddots &    &   &    &        &   &    \\
                                         &        &  1 &   &    &        &   &    \\
                                         &        &    & 1 &  0 &        &   &    \\
                                         &        &    & 0 & -1 &        &   &    \\
                                         &        &    &   &    & \ddots &   &    \\
                                         &        &    &   &    &        & 1 & 0   \\
                                         &        &    &   &    &        & 0 & -1  \end{array} \right)
\eeq

This is called the {\it normal real form} with a diagonal metric.
Here the labelling of the rows and columns is such that entries with a
minus sign correspond to the generators of the compact subalgebra
${\bf K}$, the first $r$ entries equal to $+1$ correspond to the
Cartan subalgebra, and the remaining ones to the operators in $i{\bf
P}$ {\it not} in the Cartan subalgebra.

The {\it compact real form} is obtained from the normal real form by the
Weyl unitary trick in reverse. That is, we choose the basis to be

\beq
\label{eq:compactrealform}
{\bf K}=\left\{\frac{(E_\alpha - E_{-\alpha})}{\sqrt{2}}\right\},\ \ \ \ \ \ \ 
{\bf P}=\left\{iH_i,\frac{i(E_\alpha + E_{-\alpha})}{\sqrt{2}}\right\} 
\eeq

The metric tensor is then

\beq
g_{ij}=\left( \begin{array}{cccc}  -1 &      &         &    \\
                                      & -1   &         &    \\
                                      &      &  \ddots &    \\
                                      &      &         & -1 \\
\end{array} \right)
\eeq

{\bf Example:} We will use as an example the well--known ${\bf
  SU(2,C)}$ algebra with Cartan subalgebra ${\bf H_0}=\{ J_3\}$ and
raising and lowering operators $\{ J_\pm \} $.  

\beq
\begin{array}{c}
J_3=\frac{1}{2\sqrt{2}}\tau_3,\ \ \ \ \ \ \ 
J_\pm =\frac{1}{4}(\tau_1\pm i\tau_2)\end{array}
\eeq

where 

\beq
\tau_3= \left( \begin{array}{cc} 1 & 0  \\ 
                                   0 & -1 \end{array} \right),
\ \ \ \ 
\tau_1= \left( \begin{array}{cc} 0 & 1  \\ 
                                   1 & 0  \end{array} \right),
\ \ \ \ 
\tau_2= \left( \begin{array}{cc} 0 & -i \\ 
                                   i & 0  \end{array} \right)
\eeq

We have chosen the normalization such that the non--zero entries of
$g_{ij}$ are all equal to $1$:

\beq
\begin{array}{c}
[J_3,J_\pm]=\pm \frac{1}{\sqrt{2}}J_\pm,\ \ \ \ \ \ \ 
[J_+,J_-]=\frac{1}{\sqrt{2}}J_3\end{array}
\eeq

From the commutation relations we can read off the structure constants
and determine the corresponding metric. It is non--diagonal:

\beq
g_{ij}=\left( \begin{array}{ccc} 1 & 0 & 0 \\
                                 0 & 0 & 1 \\
                                 0 & 1 & 0 \end{array} \right)
\eeq

where the rows and columns are labelled by $3,+,-$ respectively. 
To pass now to a diagonal metric, we use the recipe in eq.~(\ref{eq:NRFd})

\beq
\begin{array}{c}
\Sigma_3=J_3 \nonumber \\ \nonumber \\
\Sigma_1=\frac{J_++J_-}{\sqrt{2}}=\frac{1}{2\sqrt 2}\tau_1 \nonumber \\
\nonumber \\
\Sigma_2=\frac{J_+-J_-}{\sqrt{2}}=\frac{i}{2\sqrt 2}\tau_2 \end{array} 
\nonumber 
\eeq

The commutation relations then become

\beq
\label{eq:Sigmarels2}
\begin{array}{c} 
[\Sigma_1,\Sigma_2]=-\frac{1}{\sqrt 2}\Sigma_3,\ \ \ \ \ \ \ 
[\Sigma_2,\Sigma_3]=-\frac{1}{\sqrt 2}\Sigma_1,\ \ \ \ \ \ \ 
[\Sigma_3,\Sigma_1]=\frac{1}{\sqrt 2}\Sigma_2\end{array}
\eeq

These commutation relations characterize the algebra ${\bf
  SO(2,1;R)}$.  From here we find the non--zero structure constants
$C^3_{12}=-C^3_{21}=C^1_{23}=-C^1_{32}=-C^2_{31}=C^2_{13}=
-\frac{1}{\sqrt 2}$ and the diagonal metric of the normal real form
with rows and columns labelled $3,1,2$ (in order to comply with the
notation in eq.~(\ref{eq:NRFd_metric})) is

\beq
\label{eq:SO21}
g_{ij}=\left( \begin{array}{ccc} 1 & 0 & 0 \\
                                 0 & 1 & 0 \\
                                 0 & 0 & -1 \end{array} \right)
\eeq

which is to be compared with eq.~(\ref{eq:NRFd_metric}).  According to
eq.~(\ref{eq:NRFd}), the Cartan decomposition of ${\bf G^*}$ is ${\bf
  G^*}={\bf K}\oplus i{\bf P}$ where ${\bf K}=\{\Sigma_2\}$ and $i{\bf
  P}=\{\Sigma_3,\Sigma_1\}$.  The Cartan subalgebra consists of $\Sigma_3$.

Finally, we arrive at the compact real form by multiplying $\Sigma_3$
and $\Sigma_1$ with $i$.  Setting $i\Sigma_1=\tilde{\Sigma}_1$,
$\Sigma_2=\tilde{\Sigma}_2$, $i\Sigma_3=\tilde{\Sigma}_3$ the
commutation relations become those of the special orthogonal group:

\beq
\label{eq:tildeSigmarels2}
\begin{array}{c} 
[\tilde{\Sigma}_1,\tilde{\Sigma}_2]=-\frac{1}{\sqrt 2}\tilde{\Sigma}_3,\ \ \ \ \ \ \ 
[\tilde{\Sigma}_2,\tilde{\Sigma}_3]=-\frac{1}{\sqrt 2}\tilde{\Sigma}_1,\ \ \ \ \ \ \ 
[\tilde{\Sigma}_3,\tilde{\Sigma}_1]=-\frac{1}{\sqrt 2}\tilde{\Sigma}_2\end{array}
\eeq

The last commutation relation in eq.~(\ref{eq:Sigmarels}) has changed
sign whereas the others are unchanged.  $C^2_{31}$, $C^2_{13}$, and
consequently $g_{33}$ and $g_{11}$ change sign and we get the metric 
for ${\bf SO(3,R)}$:

\beq
\label{eq:SO3}
g_{ij}=\left( \begin{array}{ccc} -1 & 0 & 0 \\
                                 0 & -1 & 0 \\
                                 0 & 0 & -1 \end{array} \right)
\eeq 

This is a compact metric and ${\bf SO(3,R)}$ is the compact real form
of the complex algebra.  The subspaces of the compact algebra ${\bf
G}={\bf K}\oplus {\bf P}$ are ${\bf K}=\{\tilde{\Sigma}_2\}$ and ${\bf
P}=\{\tilde{\Sigma}_3,\tilde{\Sigma}_1\}$.

To summarize, real forms are obtained by using different combinations
of basis vectors (generators) and by using the Weyl unitary
trick. This changes the commutation relations and the form of the
metric. In general, a semisimple complex algebra has several distinct
real forms: one compact and several non--compact ones distinguished by
their character (trace of the metric).

\section{Obtaining all the real forms of a complex algebra} 
\label{sec-realf}
\setcounter{equation}{0}

In the previous section we saw how to construct the compact real form
of an algebra.  To classify {\it all} the real forms of any complex
Lie algebra, it suffices to enumerate all the involutive automorphisms
of its {\it compact} real form.  If ${\bf G}$ is the compact real form
of a complex semisimple Lie algebra ${\bf G^C}$, ${\bf G^*}$ runs
through all its associated non--compact real forms ${\bf G^*}$, ${\bf
G'^*}$, ... with corresponding maximal compact subgroups ${\bf K}$,
${\bf K'}$, ...  and complementary subspaces $i{\bf P}$, $i{\bf P'}$,
...  as $\sigma $ runs through all the involutive automorphisms of
${\bf G}$ ($W$ denotes the Weyl trick)

\beq
{\bf G^C} \to {\bf G}  
\begin{array}{l}{\buildrel{\scriptstyle \sigma_1 + W} \over \nearrow }\\
 {\buildrel{\scriptstyle \sigma_2 + W} \over \to } \\
 {\buildrel{\scriptstyle \sigma_3 + W} \over \searrow }
\end{array}
\begin{array}{l}
{\bf G^*}={\bf K}\oplus i{\bf P}\\ \\
{\bf G'^*}={\bf K'}\oplus i{\bf P'}\\ \\
{\bf G''^*}={\bf K''}\oplus i{\bf P''}
\end{array}
\eeq

One can show \cite{Loos} (Ch.~VII), that it suffices to consider the
following three possibilities (or combinations thereof) for $\sigma $:
$\sigma_1 =K$ (complex conjugation), $\sigma_2 = I_{p,q}$ and
$\sigma_3 = J_{p,p}$ where

\beq
\label{eq:I_J_}
I_{p,q} = \left( \begin{array}{cc} I_p & 0 \\
                                   0  & -I_q \end{array} \right), 
\ \ \ \ \ \  J_{p,p} = \left( \begin{array}{cc} 0 & I_p \\
                                   -I_p  &  0 \end{array} \right) 
\eeq

and $I_p$ denotes the $p \times p$ unit matrix.

An arbitrary involution $\sigma_i$ mixes the subspaces ${\bf K}$ and
${\bf P}$ of the compact real form and splits the algebra in a
different way into ${\bf K'}\oplus {\bf P'}$. The non--compact real
forms are then obtained through the Weyl unitary trick.

{\bf Example:}  The algebra ${\bf SO(3,R)}$, belonging to the root lattice
$B_1$ is spanned by the generators $L_1$, $L_2$, $L_3$
given in section \ref{sec-cosets}. A general element of the algebra
is

\beq
X={\bf t\cdot L}=\frac{1}{2}\left(\begin{array}{ccc}     & t^3 & t^2\\
                   -t^3 &     & t^1\\
                   -t^2 & -t^1& \end{array}\right)=
\frac{1}{2}\left(\begin{array}{ccc}     & t^3 & \\
                   -t^3 &     & \\
                        &     & \end{array}\right) \oplus
\frac{1}{2}\left(\begin{array}{ccc}     &     & t^2\\
                        &     & t^1\\
                   -t^2 & -t^1&    \end{array}\right)
\eeq

This splitting of the algebra is caused by the involution $I_{2,1}$ acting
on the representation:

\beq
I_{2,1}XI_{2,1}^{-1}=\left(\begin{array}{ccc} 1 &    &    \\
                                           &  1 &    \\
                                           &    & -1 \end{array}\right)
\frac{1}{2}\left(\begin{array}{ccc}     & t^3 & t^2\\
                                   -t^3 &     & t^1\\
                                   -t^2 & -t^1& \end{array}\right)
\left(\begin{array}{ccc} 1 &    &    \\
                           &  1 &    \\
                           &    & -1 \end{array}\right)
=\frac{1}{2}\left(\begin{array}{ccc}     & t^3 & -t^2\\
                                    -t^3 &     & -t^1\\
                                     t^2 & t^1& \end{array}\right)
\eeq

and it splits it into 
${\bf SO(3)}={\bf K}\oplus {\bf P}={\bf SO(2)}\oplus {\bf SO(3)/SO(2)}$.
Exponentiating, the coset 
representative is a point on the 2--sphere

\beq
M=\left(\begin{array}{ccc} . & .  & x\\
                         . & .  & y\\
                         . &  . & z   \end{array}\right);\ \ \ \ \ \ \  
x^2+y^2+z^2=1
\eeq

By the Weyl unitary trick we now get the non--compact real form 
${\bf G^*}={\bf K}\oplus i{\bf P}$: 
${\bf SO(2,1)}={\bf SO(2)}\oplus {\bf SO(2,1)/SO(2)}$. This algebra is
represented by 

\beq
\label{eq:SO(2,1)}
\left(\begin{array}{ccc}     & t^3 & it^2\\
                   -t^3 &     & it^1\\
                   -it^2 & -it^1& \end{array}\right)=
\left(\begin{array}{ccc}     & t^3 & \\
                   -t^3 &     & \\
                        &     & \end{array}\right) \oplus
\left(\begin{array}{ccc}     &     & it^2\\
                        &     & it^1\\
                   -it^2 & -it^1&    \end{array}\right)
\eeq

and after exponentiation of the coset generators

\beq
M=\left(\begin{array}{ccc} .  & .  & ix\\
                         .  & .  & iy\\
                         .  & .  & z   \end{array}\right);\ \ \ \ \ \ \  
(ix)^2+(iy)^2+z^2=1
\eeq

The surface in ${\bf R}^3$ consisting of points $(x,y,z)$ satisfying this
equation is the hyperboloid $H^2$. Similarly, we get the isomorphic space
$SO(1,2)/SO(2)$ by applying $I_{1,2}$: ${\bf SO(1,2)}={\bf \tilde{K}}\oplus 
i{\bf \tilde{P}}={\bf SO(2)}\oplus {\bf SO(1,2)/SO(2)}$ and in terms of the 
algebra

\beq
\tilde{X}=
\frac{1}{2}\left(\begin{array}{ccc}  &      &     \\
                                     &      & t^1 \\
                                     & -t^1 &     \end{array}\right) \oplus
\frac{1}{2}\left(\begin{array}{ccc}     & -it^3 & -it^2\\
                                  it^3  &       &      \\
                                  it^2  &       & \end{array}\right)
\eeq

\section{The classification of symmetric spaces}
\label{sec-SSclass}
\setcounter{equation}{0}

The reason we have discussed how to obtain all the real forms of a
complex algebra is that we want to understand how to obtain all the
riemannian symmetric spaces associated with the simple Lie groups.
These are namely exactly the symmetric spaces appearing as integration
manifolds in hermitean random matrix theory. (There are also
pseudo--riemannian symmetric spaces, but these are not too interesting
here.) Once we have the real forms, the symmetric spaces are defined as
the spaces corresponding to the exponential mapping of the subspaces ${\bf
P}$ and $i{\bf P}$ of the various real forms of the complex
algebra. In addition, if $G$ is a compact group, $G$ and $G^C/G$ are also 
such riemannian symmetric spaces.

To summarize, the interesting manifolds are obtained by 

\begin{itemize}
\item{obtaining the compact real form of a simple Lie algebra by combining  
ladder operators appropriately;}
\item{operating with all possible involutions on the resulting algebra,
thereby obtaining all the possible symmetric subalgebras;}
\item{forming the corresponding non--compact algebras by the Weyl unitary 
trick;}
\item{then forming pairs of symmetric coset spaces $G/K$, $G^*/K$,
$G'/K'$, $G'^*/K'$,...  corresponding to the various compact symmetric
subgroups and in addition, including the symmetric spaces $G$ and
$G^C/G$.}
\end{itemize}

\section{Curvature}
\label{sec-curv}
\setcounter{equation}{0}

A {\it curvature tensor} with components $R^i_{jkl}$ can be defined on
the manifold $G/K$ or $G^*/K$. It is given by

\beq
\label{eq:curv}
R^n_{ijk}X_n= [X_i,[X_j,X_k]] = C^n_{im}C^m_{jk}X_n
\eeq

where $\{X_i\}$ is a basis for the Lie algebra. If $\{ X,Y\} $ is an
orthonormal basis for a two--dimensional subspace $S$ of the tangent
space at a point $p$ (assumed to have dimension $\geq 2$), the {\it
sectional curvature} at the point $p$ is 

\beq
\label{eq:K}
{\cal K}=g([[X,Y],X],Y)
\eeq

For a two--dimensional manifold, this is just the Gaussian curvature.
If the manifold has dimension $\geq 2$, (\ref{eq:K}) gives the
sectional curvature along the section $S$.  These equations, together
with the commutation relations for ${\bf K}$ and ${\bf P}$, show that
the curvature of the spaces $G/K$ and $G^*/K$ has a definite and
opposite sign. To the same subgroup $K$ there corresponds a positive
curvature space $P\simeq G/K$ and a dual negative curvature space
$P^*\simeq G^*/K$.

{\bf Example:} We can use the example of $SU(2)$ to see that the
sectional curvature is the opposite for the two spaces $G/K$ and
$G^*/K$. If we take $\{ X,Y\}=\{\Sigma_3, \Sigma_1\}$ as the basis in
the space $i{\bf P}$ and $\{\tilde{\Sigma}_3, \tilde{\Sigma}_1\}$
($\tilde{\Sigma}_i\equiv i\Sigma_i$) as the basis in the space ${\bf
P}$, we see by comparing the signs of the entries of the metrics we
computed in eqs.~(\ref{eq:SO21}) and (\ref{eq:SO3}) that the sectional
curvature ${\cal K}$ at the origin has the opposite sign for the two
spaces $SO(2,1)/SO(2)$ and $SO(3)/SO(2)$.

In addition, there is also a zero--curvature symmetric space $X^0=G^0/K$
related to $X^+=G/K$ and $X^-=G^*/K$.
The space $X^0$ can be identified with
the subspace ${\bf P}$ of the algebra. The group $G^0$ is a semidirect
product of the subgroup $K$ and the invariant subspace ${\bf P}$ of
the algebra, and its elements $g=(k,a)$ act on the elements of $X^0$
in the following way:

\beq
g(x)=kx+a,\ \ \ \ k\in K,\ \ \ \  x,a \in X^0
\eeq

if the $x$'s are vectors, and 

\beq
\label{eq:matrices}
g(x)=kxk^{-1}+a,\ \ \ \ k\in K,\ \ \ \  x,a \in X^0
\eeq

if they are matrices.  The elements of the algebra ${\bf P}$ now
define an {\it abelian additive group}, and $X^0$ is a vector space
with euclidean geometry.  In the above scenario, the subspace ${\bf
P}$ contains only the operators of the Cartan subalgebra and no
others: ${\bf P}={\bf H_0}$, so that both ${\bf K}$ and ${\bf P}$ in
this case are subalgebras of ${\bf G^0}$. The algebra ${\bf G^0}={\bf
K}\oplus {\bf P}$ is non--semisimple because the subgroup ${\bf P}$ is
an abelian ideal ($[{\bf P},{\bf P}]=0$ and $[{\bf K},{\bf P}]\subset
{\bf P}$). Note that ${\bf K}$ and ${\bf P}$ still satisfy the
commutation relations in eq.~(\ref{eq:commrel}). By this equation,
$R^n_{ijk}=0$ for all the elements $X\in {\bf P}$.

Even though the Killing form on non--semisimple algebras is
degenerate, it is trivial to find a non--degenerate metric on the
symmetric space $X^0$.

{\bf Example:} An example of a flat symmetric space is $E_2/SO(2)$, where
$G^0=E_2$ is the euclidean group of motions of the plane ${\bf R^2}$:
$g(x)= kx+a$, $g=(k,a)\in G^0$ where $k\in K=SO(2)$ and $a\in {\bf
  R}^2$.  The generators of this group are translations $P_1$, $P_2
\in {\bf H_0}= {\bf P}$ and a rotation $J\in {\bf K}$ satisfying

\beq
[P_1,P_2]=0,\ \ \  [J,P_i]=-\epsilon^{ij}P_j,\ \ \ [J,J]=0 
\eeq

in agreement with eq.~(\ref{eq:commrel}) defining a symmetric
subgroup. The abelian algebra of translations $\sum_{i=1}^2t^iP_i$,
$t^i\in {\bf R}$, is isomorphic to the plane ${\bf R^2}$, and can be
identified with it.

The Killing form for $E_2$ is degenerate:

\beq
g_{ij}=\left(\begin{array}{ccc} -1 &    &    \\
                                   & 0  &    \\
                                   &    & 0  \end{array}\right)
\eeq

Therefore we don't take the Killing form as the metric on $E_2/SO(2)$. 
Instead, we can use the Euclidean metric 

\beq
\delta_{ij}=\left( \begin{array}{cc}1 & 0 \\ 0 & 1 \end{array} \right)
\eeq
 
on the entire symmetric space. 

We remark that the zero--curvature symmetric spaces correspond to the
integration manifolds of many known matrix models with physical
applications.

\section{Restricted root systems}
\label{sec-restricted}
\setcounter{equation}{0}

The restricted root systems play an important role in connection with
matrix models and integrable Calogero--Sutherland models.  Here we
will only describe very briefly how restricted root systems are
obtained and how they are related to a given symmetric space. Due to
lack of space, we will not give any example of the construction of
such a root space. A concrete example was worked out in subsection 5.2
of reference \cite{SS}.

Real forms of a complex algebra share the same root system with the
latter. This is because they correspond to the same set of raising and
lowering operators and the same Cartan subalgebra. One can also
associate a root system to a {\it symmetric space} $G/K$.  If the root
system of the group $G$ has rank $r$, the rank of this {\it restricted
root system} may be different, say $r'$.

{\bf Example:} The algebra ${\bf SU(p,q;C)}$ (${\bf p}+{\bf q}={\bf
n}$) is a non--compact real form of ${\bf SU(n,C)}$. They share the
same rank--$(n-1)$ root system $A_{n-1}$. The restricted root system
of the symmetric space $SU(p,q;C)/(SU(p)\otimes SU(q)\otimes U(1))$ is
$BC_{r'}$, where $r'={\rm min}(p,q)$.

In general the restricted root system will be different from the
original, inherited root system if the Cartan subalgebra is a subset
of ${\bf K}$. The procedure to find the restricted root system is then
to redefine the Cartan subalgebra so that it lies partly or
entirely in ${\bf P}$ (or in $i{\bf P}$, if appropriate).
This is possible if we 

\begin{itemize}

\item {find a new representation of the original Cartan subalgebra
${\bf H_0}$ corresponding to the original root lattice.  This
corresponds to a Weyl reflection of the root lattice and can be
achieved by a permutation of the root vectors. In practice it
amounts to permuting the diagonal elements of the original $H_i$'s.}

\item {do this in such a way that a {\it maximal} number $r'$ of
commuting generators are in the subspace ${\bf P}$.  The new Cartan
subalgebra ${\bf A_0}$ has the same number of generators as ${\bf
H_0}$ (this number equals the rank of the algebra), but $r'$ of its
elements lie in the subspace ${\bf P}$. $r'$ is called the rank of the
symmetric space $G/K$.}
\end{itemize} 

The new root system is defined with respect to the part of the maximal
abelian subalgebra that lies in ${\bf P}$. Therefore its rank can be 
smaller than the rank of the root system inherited from the
complex extension algebra. We can define raising and lowering operators
$E'_\alpha$ in the {\it whole} algebra ${\bf G}$ that satisfy

\beq
[X'_i,E'_\alpha]=\alpha'_i E'_\alpha \ \ \ \ \ \ \ (X'_i\in {\bf A_0}\cap
{\bf P})
\eeq

The roots $\alpha'_i$ define the restricted root system. In addition
to the sign of the curvature of the symmetric space, it is the
restricted roots (and their multiplicities) that define the Jacobian
in the transformation to radial coordinates on the symmetric space.
This Jacobian is exactly the one we encounter in random matrix theory
too, when we diagonalize the ensemble of random matrices to obtain a
partition function (and correlators) expressed only as a function of
random matrix eigenvalues. {\it The latter are exactly the radial
coordinates}. In RMT we integrate out the degrees of freedom
corresponding to the symmetric subgroup.

We have discussed the procedure for obtaining the irreducible
riemannian symmetric spaces originating in simple Lie groups. They are
listed in Table~\ref{tab1}. The classification due to E. Cartan is
based on the root systems inherited from the complex extension
algebra. We have also explained how the restricted root lattice is
defined for each symmetric space. For each compact symmetric subgroup
${\bf K}$ there is a triplet of symmetric spaces corresponding to
positive, zero, and negative curvature. The zero curvature spaces are
isomorphic to algebra subspaces ${\bf P}$ (which are the tangent
spaces of the symmetric spaces of positive curvature) and are not
listed.  The root multiplicities pertain to the {\it restricted} root
systems of the pairs of dual symmetric spaces with positive and
negative curvature.

As explained, the real forms of the simple Lie groups do not include
all the possible riemannian symmetric spaces. The compact Lie group
$G$ is itself such a space, and so is its dual $G^C/G$ (here the
algebra ${\bf G^C}={\bf G^*}\oplus i{\bf G^*}$ is the complex
extension of all the real forms ${\bf G^*}$). These are also listed
in Table~\ref{tab1}.

\begin{table}[ht]

\caption{The classification of irreducible symmetric spaces of
positive and negative curvature originating in simple Lie groups.
\label{tab1}}
\vskip5mm

\hskip-.8cm
\begin{tabular}{|l|l|l|l|l|l|l|l|}
\hline
$\begin{array}{c}Inherited\\ root\ space\end{array}$&
$\begin{array}{c}Restricted\\ root\ space\end{array}$ & $\begin{array}{c}Cartan\\ class \end{array}$ & $G/K\ (G)$ & $G^*/K\ (G^C/G)$ & $m_o$ & $m_l$ & $m_s$ \\
\hline

$A_{N-1}$ & $A_{N-1}$     & A    & $SU(N)$                 & $\frac{SL(N,C)}{SU(N)}$ & 2 & 0 & 0 \\   
          & $A_{N-1}$     & AI   & $\frac{SU(N)}{SO(N)}$   & $\frac{SL(N,R)}{SO(N)}$ & 1 & 0 & 0 \\  
          & $A_{N-1}$     & AII  & $\frac{SU(2N)}{USp(2N)}$ & $\frac{SU^*(2N)}{USp(2N)}$ & 4 & 0 & 0  \\
          &\hskip-2mm $\begin{array}{l} BC_q\ {\scriptstyle (p>q)} \\ C_q\  {\scriptstyle (p=q)} \end{array}$ 
                          & AIII & $\frac{SU(p+q)}{SU(p)\times SU(q)\times U(1)}$ & $\frac{SU(p,q)}{SU(p)\times SU(q)\times U(1)}$ & 2 & 1 & $2(p-q)$ \\
\hline

$B_N$ &$B_N$          & B    & $SO(2N+1) $               & $\frac{SO(2N+1,C)}{SO(2N+1)}$ & 2 & 0 & 2 \\

\hline

$C_N$  & $C_N$    & C    & $USp(2N)$                               &$\frac{Sp(2N,C)}{USp(2N)}$ & 2 & 2 & 0 \\
       & $C_N$    & CI   & $\frac{USp(2N)}{SU(N)\times U(1)}$      & $\frac{Sp(2N,R)}{SU(N)\times U(1)}$& 1 & 1 & 0  \\
       &\hskip-2mm $\begin{array}{l} BC_q\  {\scriptstyle (p>q)} \\  C_q\  {\scriptstyle (p=q)} \end{array}$ 
                  & CII & $\frac{USp(2p+2q)}{USp(2p)\times USp(2q)}$ & $\frac{USp(2p,2q)}{USp(2p)\times USp(2q)}$& 4 & 3 & $4(p-q)$ \\
\hline

$D_N$& $D_N$     & D          &   $SO(2N)$                            & $\frac{SO(2N,C)}{SO(2N)}$ & 2 & 0 & 0 \\
     &$C_N$     & DIII-even  & $\frac{SO(4N)}{SU(2N)\times U(1)}$     & $\frac{SO^*(4N)}{SU(2N)\times U(1)}$ & 4 & 1 & 0 \\
     &$BC_N$    & DIII-odd   & $\frac{SO(4N+2)}{SU(2N+1)\times U(1)}$ & $\frac{SO^*(4N+2)}{SU(2N+1)\times U(1)}$& 4 & 1 & 4 \\

\hline
\hskip-2mm $\begin{array}{l} B_N\ {\scriptstyle (p+q=2N+1)}\\ D_N\ {\scriptstyle (p+q=2N)}\end{array}$  
          &\hskip-2mm $\begin{array}{l} B_q\ {\scriptstyle (p>q)}\\ D_q\ {\scriptstyle (p=q)} \end{array}$
            & BDI & $\frac{SO(p+q)}{SO(p)\times SO(q)}$   & $\frac{SO(p,q)}{SO(p)\times SO(q)}$ & 1 & 0 & $p-q$ \\

\hline
\end{tabular}
\end{table}

\section{Invariant operators on symmetric spaces}
\label{sec-oper}
\setcounter{equation}{0}

Let ${\bf G}$ be a (semi)simple rank--$r$ Lie algebra. A {\it Casimir
operator} $C_k$ ($k=1,...,r$) associated with the algebra ${\bf G}$ is
a homogeneous polynomial operator that satisfies

\beq
[C_k,X_i]=0
\eeq

for all $X_i\in {\bf G}$. The simplest (quadratic) Casimir
operator associated to the adjoint representation of the 
algebra ${\bf G}$ is given by 

\beq
\label{eq:C}
C=g^{ij}X_iX_j
\eeq

where $g^{ij}$ is the inverse of the metric tensor\footnote{Note that
Casimir operators are defined for {\it semisimple} algebras, where the
Killing form is non--degenerate. This does not prevent one from
finding operators that commute with all the generators of
non--semisimple algebras.  For example, for the euclidean group $E_3$
of rotations $\{J_1,J_2,J_3\} $ and translations $\{P_1,P_2,P_3\} $,
${\bf P}^2=\sum P_iP_i$ and ${\bf P\cdot J}=\sum P_iJ_i$ commute with
all the generators. Also the operators that commute with all the
generators of a non--semisimple algebra are often referred to as
Casimir operators.}  defined in (\ref{eq:metric}) and the generators
$X_i$ are in the adjoint representation (it can be defined in a
similar way for any other representation of ${\bf G}$).

In general, the product $XY$ makes no sense in the algebra ${\bf G}$. 
The Casimir operators lie in the
enveloping algebra obtained by embedding ${\bf G}$ in the associative 
algebra defined by the relations

\beq
\label{eq:asso}
X(YZ) = (XY)Z  \ \ \ \ \ \ \ [X,Y] = XY-YX 
\eeq

The number of functionally independent Casimir operators is equal to
the rank $r$ of the group.  

All the independent Casimir operators of the algebra ${\bf G}$ can be
obtained by making the substitution $t^i\to X_i$ in the functionally
independent coefficients $\varphi_k(t^i)$ of the {\it secular equation}:

\beq
\begin{array}{c}
{\rm det}\left(\sum_{i=1}^{{\rm dim}{\bf G}} 
t^i\rho(X_i) -\lambda I_n\right)=\sum_{k=0}^n (-\lambda)^{n-k}\varphi_k(t^i)=0
\\
\varphi_k(t^i)\begin{array}{c}{\scriptstyle t^i\to X_i}\\ 
\longrightarrow \\ {} \end{array} C_l(X_i)
\end{array}
\eeq

where $\rho(X_i)$ is some representation of the algebra and 
$\varphi_k(t^i)$ are functions of the real coordinates $t^i$.  In
general there will be $r$ functionally independent coefficients, and
$r$ functionally independent Casimir operators.

{\bf Example:} The secular equation for the ${\bf SO(3)}$ rank--1 algebra is 

\beq 
{\rm det}\left({\bf t\cdot L}-\lambda I_3\right)=
\left|\begin{array}{ccc} -\lambda & t^3/2 & t^2/2 \\
                                   -t^3/2 & -\lambda & t^1/2 \\
                                   -t^2/2 & -t^1/2 & -\lambda \end{array}\right|
=(-\lambda)^3+(-\lambda)\frac{1}{4}{\bf t}^2=0
\eeq

As expected, this equation has one functionally independent
coefficient, $\varphi_1({\bf t}) =\frac{1}{4}{\bf t}^2$. The 
only Casimir operator is the square of the angular momentum
operator:

\beq
C_1\sim {\bf L}^2=L_1^2+L_2^2+L_3^2
\eeq

obtained by the substitution $t^i\to L_i$ in $\varphi_1({\bf t})$.
The Casimir operator can also be obtained from eq.~(\ref{eq:C}) by
using the metric $g_{ij}=-\frac{1}{2}\delta_{ij}$ for ${\bf SO(3)}$
given in a previous example.  We already know from quantum mechanics
that

\beq
\label{eq:L^2commrel}
[{\bf L}^2,L_1]=[{\bf L}^2,L_2]=[{\bf L}^2,L_3]=0
\eeq

The Casimir operators can be expressed as differential
operators in the local coordinates on the symmetric space:

\beq
X=\sum_\alpha X^\alpha(x) \partial_\alpha \equiv \sum_\alpha X^\alpha(x) \frac{\partial }{\partial x^\alpha }
\eeq

where $x^\alpha $ are local coordinates \cite{Helgason,SattW} (for
example, $L_x=({\bf r \times p})_x=-i(y\partial_z-z\partial_y)$).

Expressed in local coordinates as differential operators, the Casimirs
are called {\it Laplace operators}. In analogy with the Laplacian in
${\bf R^n}$,

\beq
{\bf P}^2=\Delta=\sum_{i=1}^n\frac{\partial^2}{\partial {x^i}^2}
\eeq

which is is invariant under the group $E_n$ of rigid motions
(isometries) of ${\bf R^n}$, the Laplace operators on
(pseudo--)riemannian manifolds are invariant under the group of
isometries of the manifold.  The isometry group of the symmetric space
$P\simeq G/K$ is $G$, since $G$ acts transitively on this space and
preserves the metric.  The number of independent Laplace operators on
a riemannian symmetric coset space is equal to the rank of the space.

The {\it Laplace--Beltrami operator} on a symmetric space is the
special second order Laplace operator. It can be expressed as

\beq
\label{eq:L-B-op}
\Delta_Bf
=\frac{1}{\sqrt{|g|}}\frac{\partial }{\partial x^i} g^{ij}
\sqrt{|g|}\frac{\partial }{\partial x^j}f,\ \ \ \ \ \ \ 
g\equiv {\rm det}g_{ij}
\eeq

{\bf Example:} Let's calculate the Laplace--Beltrami operator on the
symmetric space $SO(3)/SO(2)$ in polar coordinates using
(\ref{eq:L-B-op}) and the metric at the point $(\theta,\phi)$ 

\beq
g_{ij}=\left(\begin{array}{cc} 1 & 0 \\
                       0 & {\rm sin}^2\theta \end{array}\right),\ \ \ \ \ \ \  
g^{ij}=\left(\begin{array}{cc} 1 & 0 \\
                       0 & {\rm sin}^{-2}\theta \end{array}\right) 
\eeq

Substituting in the formula and computing derivatives we obtain the 
Laplace--Beltrami operator on the sphere of radius $1$:

\beq
\label{eq:Delta_on_sphere}
\Delta_B=\partial_\theta^2+{\rm cot}\theta\, \partial_\theta+
{\rm sin}^{-2}\theta \, \partial_\phi^2
\eeq

Of course this operator is proportional to ${\bf L}^2$. We can
check this by computing $L_x= -i(y\partial_z-z\partial_y)$, $L_y=
-i(z\partial_x-x\partial_z)$, and $L_z= -i(x\partial_y-y\partial_x)$
in spherical coordinates (setting $r=1$) and then forming the operator
$L_x^2+L_y^2+L_z^2$, remembering that all the operators have to act
also on anything coming after the expression for each $L_i^2$. We find
that ${\bf L}^2$ in spherical coordinates, expressed as a differential
operator, is exactly the Laplace--Beltrami operator.

As we have seen in eq.~(\ref{eq:sphdec}), radial coordinates can be
defined on the SS.  The adjoint representation of a general element
$H$ in the maximal abelian subalgebra ${\bf H_0'}\subset {\bf P}$
follows from a form similar to eq.~(\ref{eq:adjrad}) (with or without
a factor of $i$ depending on whether we have a compact or non--compact
space), but now the roots are in the {\it restricted} root lattice.
For a non--compact space of type $P^*$

\beq
\label{eq:q^alpha}
{\rm log}h=H={\bf q\cdot H}=\left(\begin{array}{cccccc}
0 & & & & & \\ & \ddots & & & & \\ & & 0 & & & \\ & & & {\bf q\cdot \alpha } & & \\
 & & & & \ddots & \\ & & & & & -{\bf q\cdot \eta }  \end{array}\right) 
\eeq

We define $q^\alpha \equiv{\bf q\cdot \alpha }$. These are the {\it
radial coordinates on the symmetric space}.

{\bf Example:} The rank of the symmetric space $SU(2,C)/SO(2)$ is 1
and the restricted root lattice is $A_1$. Absorbing a factor of
$\sqrt{2}$ (the length of the ordinary roots) into the coordinate, the
above equation takes the form

\beq
H=\theta H_1=\theta \left(\begin{array}{ccc}
0 & &  \\  & 1 & \\ & & -1 \end{array}\right),
\ \ \ \ \ \ \ 
h={\rm e}^{i\theta H_1}=\left(\begin{array}{ccc}
1 & &  \\  & {\rm e}^{i\theta } & \\ & & {\rm e}^{-i\theta }\end{array}\right)
\eeq

The radial coordinate is $q=(q^1)=\theta $.

{\bf Example:} From Table \ref{tab1}, on the symmetric negative
curvature space $SO(2N,C)/SO(2N)$, the restricted root lattice is of
type $D_N$ and the roots are $\{ \pm e_i \pm e_j, i\neq j \}$. The
$q^\alpha $ are $\pm q_i \pm q_j$.

In general, a Laplace--Beltrami operator can be split into a radial
part $\Delta_B'$ and a transversal part. The radial part acts on
geodesics orthogonal to some submanifold $S$, typically a sphere
centered at the origin \cite{Helgason2}.

{\bf Example:} The radial part of the Laplace--Beltrami operator
for the coset space $SO(3)/SO(2)$ given in (\ref{eq:Delta_on_sphere}) is

\beq
\label{eq:Delta'_on_sphere}
\Delta'_B=\partial_\theta^2+{\rm cot}\theta\, \partial_\theta
\eeq

The radial part of the Laplace--Beltrami operator on a symmetric space
has the general form

\beq
\label{eq:DeltaB'} 
\Delta_B'= \frac{1}{J^{(j)}}\sum_{\alpha =1}^{r'}\frac{\partial }
{\partial q^\alpha }J^{(j)}\frac{\partial }{\partial q^\alpha }\ \ \ \ \ \ \ 
(j=0,-,+)
\eeq

where $r'$ is the rank of the symmetric space, $J^{(j)}$ is the
Jacobian of the transformation to radial coordinates on the SS (to be
given below) and $m_\alpha $ is the multiplicity of the restricted
root $\alpha $. (The multiplicities $m_\alpha $ were listed in
Table~\ref{tab1}.)  
The sum in (\ref{eq:DeltaB'}) goes over 
the labels of the independent radial
coordinates $q={\rm log}h(x)= (q^1,...,q^{r'})$ where $h(x)$ is the
exponential map of an element in the Cartan subalgebra.

The Jacobian in (\ref{eq:DeltaB'}) is given by

\beq
\label{eq:J_j}
\begin{array}{l}
J^{(0)}(q)=\prod_{\alpha \in R^+} (q^\alpha )^{m_\alpha }\\
\\
J^{(-)}(q)=\prod_{\alpha \in R^+} ({\rm sinh}(q^\alpha ))^{m_\alpha }\\ 
\\
J^{(+)}(q)=\prod_{\alpha \in R^+} ({\rm sin}(q^\alpha ))^{m_\alpha }\end{array}
\eeq

for the various types of symmetric spaces with zero, negative and
positive curvature, respectively (see \cite{Helgason2}, Ch.~I,
par.~5).  $J^{(j)}=\sqrt{|g|}$ where $g$ is the metric tensor at an
arbitrary point of the symmetric space.  In these equations the
products denoted $\prod_{\alpha \in R^+}$ are over all the positive
roots of the restricted root lattice.\footnote{Strictly speaking, in
the euclidean case we have not defined any restricted root lattice.
The formula for the Jacobian $J^{(0)}(q)$ for the zero--curvature
space is understood as the infinitesimal version of the formula
pertaining to the negative--curvature space.}.

{\bf Example:} If the restricted root lattice is of type $A_N$ with
only ordinary roots $e_i-e_j$ ($i\neq j$), the Jacobian of the zero
curvature space is

\beq
\label{eq:gaussJ'}
J^{(0)} (\{q_i\})= \prod_{i<j} |\,q_i-q_j\,|^{m_o} \nonumber
\eeq

The absolute value corresponds to a certain choice of Weyl chamber
(ordering of the $q_i$) \cite{SS}.

{\bf Example:} If the restricted root lattice is of type $C_N$ with 
long and ordinary roots, the positive roots are $\{e_i\pm e_j,2e_i\}$. 
The Jacobian of the negative curvature space is then 

\beq
\label{eq:transferJ''}
J^{(-)} (\{q_i\})= \prod_{i<j} |\,{\rm sinh}^2q_i-
{\rm sinh}^2q_j\,|^{m_o} \prod_{k}{\rm sinh}^{m_l}(2q_k)
\eeq

where we have used ${\rm sinh}(q_i-q_j){\rm sinh}(q_i+q_j)={\rm
sinh}^2q_i- {\rm sinh}^2q_j$. (The root multiplicities were listed in
Table~\ref{tab1}.)

Since the Laplace operators form a commutative algebra, they have
common eigenfunctions.  The eigenfunctions of the radial part of the
Laplace--Beltrami operator on a symmetric space are called {\it zonal
spherical functions}. They play an important role in mathematics as
bases for square--integrable functions, not to mention their role as
irreducible representation functions in quantum mechanics.  In the
present context we will see that they determine the solution of some
physical problems where the relevant operator can be mapped onto the
Laplace--Beltrami operator.

Suppose the smooth complex--valued function $\phi_\lambda (x)$ is an
eigenfunction of some invariant differential operator $\Delta_k$ on
the symmetric space $G/K$:

\beq
\Delta_k \phi_\lambda (x)=\gamma_{\Delta_k}(\lambda ) \phi_\lambda (x)
\eeq

The function $\phi_\lambda (x)$ is called spherical if it satisfies
$\phi_\lambda (kxk') = \phi_\lambda (x)$ ($x\in G/K$, $k\in K$) and if
$\phi_\lambda (e)=1$ ($e=$identity element).  Because of the
bi--invariance under $K$, these functions depend only on the radial
coordinates $h$:

\beq
\phi_\lambda (x)=\phi_\lambda (h)
\eeq

{\bf Example:} We know from quantum mechanics that the eigenfunctions
of the Laplace operator ${\bf L}^2$ on $G/K=SO(3)/SO(2)$ are the
associated Legendre polynomials $P_l({\rm cos}\theta )$. Setting ${\bf
L}_x=-i(y\partial_z-z\partial_y)$ etc.,

\beq
\label{eq:eigenPl}
{\bf L}^2P_l({\rm cos}\theta )=l(l+1)P_l({\rm cos}\theta )
\eeq

where ${\rm cos}\theta $ is the $z$--coordinate of the point
$P=(x,y,z)$ on the sphere of radius $1$ (in spherical coordinates,
$P=({\rm sin}\theta \, {\rm cos}\phi , {\rm sin}\theta \, {\rm
sin}\phi , {\rm cos}\theta )$).  As we can see, the eigenfunctions are
functions of the radial coordinate $\theta $ only. The subgroup that
keeps the north pole fixed is $K=SO(2)$ and its algebra contains the
operator $L_z=\partial_\phi $. Indeed, $P_l({\rm cos}\theta )$ is
unchanged if the point $P$ is rotated around the $z$--axis.

Following reference \cite{OlshPere}, we introduce a parameter $a$ 
into the the Jacobians (\ref{eq:J_j}) for the symmetric spaces,

\beq
\label{eq:J_ja}
\begin{array}{l}
J^{(0)}(q)=\prod_{\alpha \in R^+} (q^\alpha )^{m_\alpha }\\
\\
J^{(-)}(q)=\prod_{\alpha \in R^+} (a^{-1}{\rm sinh}(aq^\alpha ))^{m_\alpha }\\ 
\\
J^{(+)}(q)=\prod_{\alpha \in R^+} (a^{-1}{\rm sin}(aq^\alpha ))^{m_\alpha }\end{array}
\eeq

The parameter $a$ corresponds to a radius. For example, for the sphere
$SO(3)/SO(2)$ it is the radius of the 2--sphere.

The various spherical functions corresponding to the spaces of
positive, negative and zero curvature are then related to each other
by the simple transformations \cite{OlshPere}

\beq
\label{eq:-to0+}
\begin{array}{l}
\phi_\lambda^{(0)} (q)={\rm lim}_{a\to 0} \phi_\lambda^{(-)}(q) \\   \\
\phi_\lambda^{(+)} (q)=\phi_\lambda^{(-)}(q)|_{a\to ia} \end{array}
\eeq   

and their eigenvalues are given by 

\beq
\label{eq:eigen}
\begin{array}{l}
\Delta_B'\phi_\lambda^{(0)}= -\lambda^2 \phi_\lambda^{(0)}\\ \\
\Delta_B'\phi_\lambda^{(-)}= (-\frac{\lambda^2}{a^2}-\rho^2)\phi_\lambda^{(-)}\\ \\
\Delta_B'\phi_\lambda^{(+)}= (-\frac{\lambda^2}{a^2}+\rho^2)\phi_\lambda^{(+)}
\end{array}
\eeq

where $\rho $ is the function defined by

\beq
\label{eq:rho}
\rho=\frac{1}{2} \sum_{\alpha \in R^+}m_\alpha \alpha
\eeq

There is an extensive theory relating to such eigenfunctions
\cite{Helgason2}, but we will not be able to discuss it here. Some
important results and references were listed in \cite{SS}.

\section{A new classification of RMT}
\label{correspond}
\setcounter{equation}{0}

In this third lecture we will discuss three applications of the theory
of symmetric spaces in random matrix theory.  We have not been able to
dwell upon the details of all the various random matrix
ensembles. However, they all have some common features that we have
already discussed in the introductory part of these lectures:

\begin{itemize} 

\item{the ensembles are determined by physical symmetries and labelled
by a Dyson index $\beta $ counting the degrees of freedom of the
matrix elements (there is also a boundary index $\alpha $
characterising the ensembles that do not have translationally 
symmetric eigenvalue distributions);}

\item{the probability $P(M)dM$ is invariant under some similarity
transformation of the matrix $M$;}

\item{the random matrix integral can be expressed as a function of
random matrix eigenvalues by diagonalizing the ensemble;}

\item{the Jacobian due to this diagonalization determines the geometric
eigenvalue repulsion characteristic of RMT;}

\item{one can identify ensembles of random matrices with symmetric spaces.}
\end{itemize}

We will now discuss the correspondence between matrix ensembles and
symmetric spaces in more detail through a few examples.

\begin{figure}
\centerline{
\psfig{figure=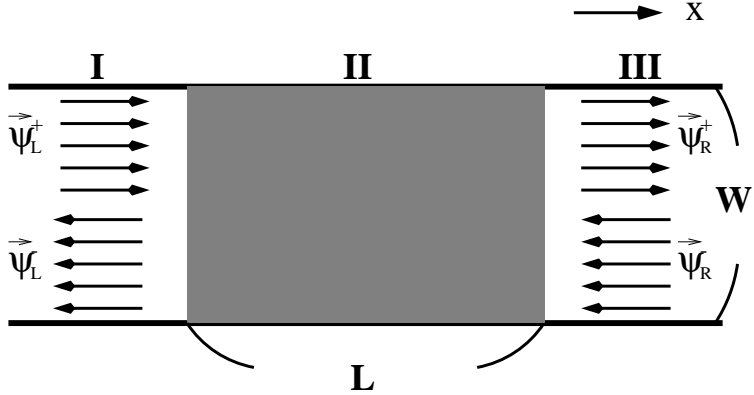,width= 10cm}
}%
\caption[]{A schematic picture of a quantum wire consisting of a
disordered region (II) of length $L$ and width $W$ connected to
electron reservoirs through leads (I, III). In each lead there are $N$
incoming and $N$ outgoing scattering channels.}
\label{fig_qwire}
\end{figure}

{\bf Example:} The circular random matrix ensembles are ensembles of
random scattering matrices $S$ relating the incoming and the outgoing
wave amplitudes in a scattering problem. Scattering processes are
important both in mesoscopic physics and in many--body problems in
nuclear and atomic physics.  The scattering system is idealized as
incoming and outgoing scattering channels in which propagation is
free, and a compact interaction region where the scattering takes
place.  Let us for simplicity consider scattering in a mesoscopic
disordered system connected to electron reservoirs through leads.  In
Fig.~\ref{fig_qwire} a schematic wire of length $L$ and width $W$ is
shown (the region II is disordered).  The wave functions of incoming
and outgoing electrons in the left and right leads are denoted
$\Psi^{\pm}_{R,L}(\vec{r})$. A wave function $\Psi^{\pm}(\vec{r})$ can
be decomposed as

\beq
\Psi^{\pm}_n(\vec{r})=\Phi_n(\vec{r}_t){\rm e}^{\pm ik_nx}
\eeq
 
where $\Phi_n(\vec{r}_t)$ is the transverse wave function and the
integer $n=1,...,N$ labels the $N$ propagating {\it modes} or {\it
scattering channels}.  The coordinate $x$ is along the wire.  

The scattering matrix $S$ relates the incoming and the outgoing wave
amplitudes of the electrons. If we have $N$ propagating modes
at the Fermi level, we can describe them by a vector of
length $2N$ of incident modes $I$, $I'$ and a similar vector of
outgoing modes $O$, $O'$ in each lead, where unprimed letters denote
the modes in the left lead and primed letters the modes in the right
lead. Then the scattering matrix is defined by

\beq
\label{eq:scat}
S\left( \begin{array}{c} I \\ I' \end{array} \right) =
\left( \begin{array}{c} O \\ O' \end{array} \right)
\eeq

Flux conservation ($|I|^2+|I'|^2=|O|^2+|O'|^2$), implies that $S$ is
unitary. The symmetry classes we discussed in connection with Wigner's
hamiltonian ensembles are reflected here in the circular orthogonal
ensemble ($\beta=1$, $S$ is unitary and symmetric), the cirular
unitary ensemble ($\beta=2$, $S$ is unitary) and the circular
symplectic ensemble ($\beta=4$, $S$ is unitary and self-dual). (Note
that the hermitean Hamiltonians can be related to the unitary
scattering matrix by $S=\e^{iH}$.)

Let's look more closely at the circular orthogonal ensemble. Every
symmetric unitary matrix $S$ can be written as

\eq
S=U^TU
\label{dy1}
\en

where $U$ is a generic unitary matrix. However, this mapping is not
one--to--one. If we assume that $S=U^TU=V^TV$, then it is easy to see
that the matrix relating the two expressions, $R=VU^{-1}$, is unitary
and satisfies $R^TR=1$ \cite{Dyson}. Hence $R$ must be real and
orthogonal.  Thus we see that the manifold of the unitary symmetric
matrices is actually the coset $U(N)/O(N)$, due to the above mentioned
degeneracy. From the point of view of the physical properties of the
ensemble nothing changes if we perform the restriction to an
irreducible symmetric space.\footnote{In the partition function

\beq
Z\sim \int_{G/K} dS\, P_{\beta }(S)
\eeq

extracting such a $U(1)$ factor from the integration manifold just
amounts to redefining $Z$ by a constant.} Then the manifold becomes
$SU(N)/SO(N)$. The integration manifold in the unitary ensemble is
simply the group $SU(N)$ without further constraint and in the
symplectic ensemble, in an analogous manner to the orthogonal case, we
realize that the manifold coincides with $SU(2N)/Sp(2N)$.  Comparing
now with Table~\ref{tab1}, we see that the integration manifolds of the
three circular ensembles are exactly the first three coset spaces
(described in the Cartan notation as A, AI and AII) of positive
curvature in the list of possible irreducible symmetric spaces.

{\bf Example:}  As implied in the discussion of hamiltonian ensembles
in section \ref{sec-symmetries}, $P_\beta (H)$ and the integration
measure $dH$ are separately invariant under the transformation

\beq
\label{eq:UHU}
H \to UHU^{-1},
\eeq

where $U$ is an orthogonal, unitary or symplectic $N\times N$ matrix,
depending on the value of $\beta $. It can be shown \cite{Mehta}
that the form of $P_\beta (H)$ is automatically restricted to the form

\beq
P_\beta (H)={\rm exp}(-a\, {\rm tr}H^2+b\, {\rm tr}H+c)
\eeq

($a>0$) if one postulates statistical independence of the matrix
elements $H_{ij}$.   Note that $P_\beta (H)$ can be cast in the form

\beq
\label{eq4.1}
P_\beta (H) \sim \e^{-a \, \tr H^2}
\eeq

by simply completing the square in the exponent. As we know, this is a
good choice for the RMT potential, because it makes the theory easy to
solve.

However, it is important to notice that the symmetry group of $P_\beta
(H)\,dH$ is larger and consists of rotations by the matrix $U$ like in
eq.~(\ref{eq:UHU}), and addition by square hermitean matrices:

\beq
H\to UHU^{-1}+H' 
\label{UHUH}
\eeq

This latter equation tells us that the ensemble is translation
invariant.  Limiting the discussion again to irreducible symmetric
spaces, for $\beta=1$ we are then dealing with the set of real,
symmetric and traceless matrices.  But this is exactly the {\it
algebra subspace} ${\bf SL(N,R)/SO(N)}$ corresponding to a symmetric
space of euclidean type (zero curvature) (cf. section \ref{sec-ss})
and obtained by removing the set of real, {\it antisymmetric} and
traceless matrices from the algebra ${\bf SL(N,R)}$. 

By performing a similar analysis for $\beta=$2, 4 we obtain the
following general result: The gaussian ensembles labelled by $\beta
$=2, 4 consist of hermitean square matrices belonging to algebra
subspaces ${\bf SL(N,C)/SU(N)}$ and ${\bf SU^*(2N)/USp(2N)}$,
respectively.  From Table~\ref{tab1} we see that these three symmetric
spaces correspond to algebra subspaces in Cartan classes A, AI and
AII. The integration manifolds of the circular ensembles are the
positive curvature coset spaces corresponding to the same Cartan
classes.

Also the chiral ensembles used in field theory correspond to algebra
subspaces. They are identified respectively with ${\bf
SO(p,q)/(SO(p)\otimes SO(q))}$, ${\bf SU(p,q)/(SU(p)\otimes SU(q))}$,
or ${\bf USp(p,q)/(USp(p)\otimes USp(q))}$ (in this case ${\bf p}$,
${\bf q}$ have to be even). We will not discuss them further here (for
details we refer to \cite{SS}).

{\bf Example:} The transfer matrix ensembles appear in the theory of quantum
transport, in the random matrix theory description of so called
quantum wires.  In these pages we will only discuss the part of the
theory which is relevant for our purpose, the study of the mapping
between random matrix theory and symmetric spaces.

The natural theoretical framework for describing mesoscopic systems is
the Landauer theory~\cite{lan}. Within this approach Fisher and Lee
\cite{fl} proposed the following expression for the conductance in a
two--probe geometry (a finite disordered section of wire to which
current is supplied by two semi-infinite ordered leads):

\eq
\label{eq:1a}
G=G_0~{\rm Tr}(tt^\dagger)\equiv G_0\sum_n T_n,~~~~~~~
G_0=\frac{2e^2}{h}
\en

where $t$ is the $N\times N$ transmission matrix of the conductor (see
eq.~(\ref{eq:Smatrix}) below), $N$ is the number of scattering channels at
the Fermi level and $T_1,T_2 \cdots T_N$ are the eigenvalues of the
matrix $tt^\dagger$. The $T_i$'s are usually referred to as transmission
eigenvalues.  The constants $e$ and $h$ denote the electronic charge
and Planck's constant, respectively.

While the scattering matrix $S$ relates the incoming wave
amplitudes $I$, $I'$ to the outgoing wave amplitudes $O$, $O'$ (see
eq.~(\ref{eq:scat})), the transfer matrix $M$ relates the wave
amplitudes in the left lead to those in the right lead:

\beq
M\left( \begin{array}{c} I \\ O \end{array} \right) =
\left( \begin{array}{c} O' \\ I' \end{array} \right)
\eeq

The transfer matrix formalism is more appropriate for description of
$1d$ systems than the scattering matrix formalism. This is due to the
multiplicative property of the transfer matrix, as an infinitesimal
slice is added to the quantum wire.

Following a standard notation \cite{BILP,MPK} the scattering matrix
has the following block structure

\beq
\label{eq:Smatrix}
S=\left( \begin{array}{cc} r & t' \\
                          t & r' \end{array} \right)
\eeq

where $r$, $r'$, $t$, $t'$ are $N\times N$ reflection and transmission
matrices.  The unitarity of $S$ implies that the four matrices
$tt^{\dagger }$, $t't'^{\dagger }$, $1-rr^{\dagger }$,
$1-r'r'^{\dagger }$ have the same set of eigenvalues $T_1,...,T_N$ ($0
\leq T_i \leq 1$).  The parameters\footnote{The $\lambda_i$ are the
eigenvalues of the matrix

\beq
\label{eq:Qlam}
Q=\frac{1}{4}(M^{\dagger}M +(M^{\dagger}M)^{-1}-2)
\eeq
} $\lambda_i$ are related to the transmission eigenvalues by

\beq
\label{eq:Tlambda}
\lambda_i=\frac{1-T_i}{T_i}
\eeq
 
The $\lambda_i$ are non--negative. In terms of these, $M$ can be parametrized 
as \cite{MPK}

\beq
\label{eq:Mpara}
M=\left( \begin{array}{cc} u & 0 \\
                           0 & u' \end{array} \right) 
\left( \begin{array}{cc} \sqrt{1+\Lambda} & \sqrt{\Lambda} \\
                         \sqrt{\Lambda} & \sqrt{1+\Lambda}\end{array} \right) 
\left( \begin{array}{cc} v & 0 \\
                         0 & v' \end{array} \right) \equiv U\Gamma V
\eeq

where $u$, $u'$, $v$, $v'$ are unitary $N \times N$ matrices (related
by complex conjugation: $u'=u^*$, $v'=v^*$ if $M\in Sp(2N,R)$ or $M\in
SO^*(4N)$, see below) and $\Lambda = {\rm
  diag}(\lambda_1,...,\lambda_N)$.  In case spin--rotation symmetry is
broken, the number of degrees of freedom in (\ref{eq:Mpara}) is
doubled and the matrix elements are real quaternions.

Transfer matrices are strongly constrained by various physical
requirements.  Flux conservation, presence or absence of
time--reversal symmetry, and presence or absence of spin--rotation
symmetry lead to conditions on the transfer matrix. These conditions
determine the group $G$ to which $M$ belongs. For example, flux
conservation leads to the following condition on $M$ \cite{MPK}

\beq
\label{eq:fluxcons}
M^{\dagger }\Sigma_z M =\Sigma_z, \ \ \ \ \ \ \Sigma_z=\left( \begin{array}{cc}
1 & 0 \\ 0 & -1 \end{array} \right)
\eeq

i.e. $M$ preserves the ($2N\times 2N$) metric $\Sigma_z$. This means
that $M$ belongs to the pseudo--unitary group $SU(N,N)$ ($M$ has to be
continuously connected to the unit matrix so we take the connected
component of $U(N,N)$). For a detailed discussion of all the above constraints
we refer the reader to \cite{SS} and references therein.

Using the parametrization in eq.~(\ref{eq:Mpara}) it is easy to check
that rotating $M$ by a matrix $W\in SU(N) \times SU(N) \times U(1)$
gives a new transfer matrix $M'=WMW^{-1}=U'\Gamma V'$ with the {\it
same} matrix $\Gamma $ and therefore the same physical degrees of
freedom $\{\lambda_1,...,\lambda_N\}$.  This means that the matrix
$\Gamma $ belongs to a coset space $G/K$, where $M\in G$ and $W\in
K$. Following in each case a similar reasoning, one obtains three
ensembles of transfer matrices corresponding to different physical
symmetries: $Sp(2N,R)/U(N)$, $SU(N,N)/SU(N) \times SU(N) \times U(1)$
and $SO^*(4N)/U(2N)$.  These are symmetric spaces of negative
curvature, as is also evident from Table~\ref{tab1}.  They correspond
to Cartan classes CI, AIII, and DIII--even, respectively.

In the three examples above, it was evident that the physical degrees
of freedom in RMT correspond to points in the symmetric coset spaces
in Cartan's classification. To every symmetric space a random matrix
ensemble is associated, and many of these have been directly applied
in some physical context (see \cite{SS} for more details). The
ensembles listed in Table~\ref{tab2} are examples of such RMT's.

In section \ref{sec-restricted} we identified the random matrix
eigenvalues with the radial coordinates on the symmetric space. Their
probability distribution is determined by the Jacobian of the
transformation to spherical coordinates. This Jacobian was given in
eq.~(\ref{eq:J_j}).  It its determined by the curvature of the
underlying symmetric space, and in addition by the roots and root
multiplicities $m_\alpha $ of the restricted root lattice
(cf. eq.~(\ref{eq:q^alpha})). As a consequence, {\it the form and
strength of the characteristic eigenvalue repulsion in RMT is
determined by the root lattice of the underlying symmetric space!}
What's more, in the following examples we will also identify the Dyson
and boundary indices of the random matrix ensembles as determined by
these multiplicities, and we will see that the $m_\alpha $ determine
the classical orthogonal polynomials related to each matrix model.

{\bf Example:} The Jacobian associated to the chiral random matrix ensembles 

\beq
J_{\beta }^{\nu }(\{\lambda_i\})\propto \prod_{i<j}
|\lambda_i^2-\lambda_j^2|^\beta \prod_{k}|\lambda_k|^{\beta (\nu +1)-1}
\eeq

(where $\nu $ is the number of zero modes) corresponds in the most
general case to a root lattice with all types of roots.  The chiral
ensembles are algebra subspaces (this can be deduced also from the
block--structure of chiral matrices). Like for the gaussian ensembles
they are associated with the restricted root lattice of the curved
symmetric spaces in the same triplet (since we have not defined any
such root lattice for flat symmetric spaces). The restricted root
lattices for the chiral ensembles are of type $B_q$ or $D_q$ for
$\beta =1$ and $BC_q$ or $C_q$ for $\beta =2,4$.  The positive roots
are as follows: for $B_q$ $\{ e_i, e_i\pm e_j\}$, for $D_q$ $\{ e_i\pm
e_j\}$, for $BC_q$ $\{ e_i, e_i\pm e_j, 2e_i\}$, for $C_q$ $\{ e_i\pm
e_j,2e_i\}$ ($i\neq j$ always). Using the root multiplicities
$m_o=\beta $, $m_l=\beta -1$, $m_s=\beta |p-q|\equiv \beta \nu $
(cf. the values in Table~\ref{tab1}) we see that the Jacobian is of
type (\ref{eq:J_ja}).  It can be rewritten

\beq
\label{eq:J0}
J^{(0)}(\{\lambda_i\})\sim \prod_{i<j} |\lambda_i^2-\lambda_j^2|^{\beta }
\prod_k |\lambda_k|^\alpha,\ \ \ \ \ \ \ \beta \equiv m_o,\ \ \alpha \equiv m_s+m_l
\eeq

From this Jacobian it is evident that in addition to the usual
repulsion between different eigenvalues, $\lambda_i$ also repels its
mirror image $-\lambda_i$, and the eigenvalues are no longer
translationally invariant. This kind of ensembles are therefore called
boundary random matrix theories and $\alpha$ is called a boundary
index.  The boundary random matrix theories (BRMT) include chiral and
normal--superconducting (NS) ensembles.  We have given just one
example, but one can make the same identifications in all the RMT's.

Known symmetries of the random matrix ensembles can be understood
in terms of the symmetries of the associated restricted root
lattice. In particular, ensembles of $A_n$ type are characterized by
translational invariance of the eigenvalues. This translational
symmetry is seen to originate in the root lattice: all the restricted
roots of the $A_n$ lattice are of the form $(e_i-e_j)$.
For all the other types of restricted root lattices ($B_n$, $C_n$,
$D_n$ and $BC_n$) this invariance is broken and substituted by a new
$Z_2$ symmetry giving rise to the reflection symmetry of the
eigenvalues discussed above. 

As we saw in the introduction to random matrix ensembles, orthogonal
polynomials are an important tool for the exact calculation of
eigenvalue correlation functions in RMT.  Interesting in our framework
is that the parameters which define the polynomials can be explicitly
related to the multiplicities of short and long roots of the
underlying symmetric space and thus, by the identification in
Table~\ref{tab2}, with the boundary universality indices of the BRMT.
The relations are the following:

{\bf Laguerre polynomials:}

\bea
L^{(\lambda)}(x)=\frac{x^{-\lambda}e^x}{n!}\frac{d^n}{dx^n}
(x^{n+\lambda} e^{-x})\ \ \ \ \ (x\geq 0) \nonumber \\
\lambda \equiv \frac{m_s+m_l-1}{2} \\ \nonumber
\eea

{\bf Jacobi polynomials:}

\bea
P^{(\rho,\sigma )}(x)=\frac{(-1)^n}{2^nn!}
\frac{(1-x)^{-\sigma }}{(1+x)^\rho }\frac{d^n}{dx^n}
\left( \frac{(1+x)^{n+\rho }}{(1-x)^{-n-\sigma }} \right) \ \ \ \ \ 
(-1\leq x\leq 1)
\nonumber \\
\rho \equiv \frac{m_s+m_l-1}{2},\ \ \sigma \equiv \frac{m_l-1}{2}\\ \nonumber
\eea

We see that $\lambda $ and $\rho $ have the same expression in terms
of $m_s$ and $m_l$. Thus the BRMT's corresponding to Laguerre and
Jacobi polynomials with the same $\lambda =\rho $ indices belong to
the same triplet in the classification of Table~\ref{tab3a}. They are
respectively the zero curvature (Laguerre) and positive curvature
(Jacobi) elements of the triplet. This explains the fact that 
scaled correlators are the same for Laguerre and Jacobi ensembles
of the same $\beta $ near the boundary (so called ``weak
universality'') \cite{for}.

As a consequence of the identification of symmetric spaces and random
matrix ensembles, the physical systems corresponding to random matrix
ensembles can be organized into universality classes. In
Table~\ref{tab3a} random matrix ensembles with known physical
applications are listed in the columns labelled $X^+$, $X^0$ and $X^-$
and correspond to symmetric spaces of positive, zero and negative
curvature, respectively.  Extending the notation used in the
applications of chiral random matrices in QCD, where $\nu $ is the
winding number, we set $\nu \equiv p-q$.  The notation is C for
circular, G for gaussian\footnote{Strictly speaking, these two groups
are the scattering matrix and hamiltonian Wigner--Dyson ensembles, of
which the latter have come to be referred to as ``gaussian ensembles''
due to the most common choice of random matrix potential, and the
former are called circular because the eigenvalues of the unitary
matrices lie on the unit circle.}, $\chi $ for chiral, B for
Bogoliubov--de Gennes, P for $p$--wave, T for transfer matrix and S
for S--matrix ensembles\footnote{For a more detailed discussion of all
these ensembles and their relation to symmetric spaces we refer to
\cite{SS} and references therein.}.  The upper indices indicate the
curvature, while the lower indices correspond to the multiplicities of
the restricted roots characterizing the spaces with non--zero
curvature. To the euclidean type spaces $X^0\sim G^0/K$, where the
non--semisimple group $G^0$ is the semidirect product $K\otimes {\bf
P}$, we associate the root multiplicities of the algebra ${\bf G}={\bf
K}\oplus {\bf P}$. Note that the ensembles that have been given are
the ones to which we have found explicit reference in the literature
(see \cite{SS}). In principle all the empty boxes could be filled too.

In Table \ref{tab2} we list some identifications made between matrix
ensembles and symmetric spaces. We have not discussed the Coulomb gas
analogy, but most of the other entries we have or will mention. For a
more thorough discussion and references we refer to \cite{SS}.

\begin{table}
\caption{Irreducible symmetric spaces and some of their random
matrix theory realizations. 
\label{tab3a}}
\vskip5mm

\hskip-1cm
\begin{tabular}{|l|l|l|l|l|l|l|l|l|l|}
\hline
$\begin{array}{c}Restricted\\ root\ space\end{array}$ & $\begin{array}{c}Cartan\\ class \end{array}$ & $G/K\ (G)$ & $G^*/K\ (G^C/G)$ & $m_o$ & $m_l$ & $m_s$ & $X^+$ & $X^0$ & $X^-$\\
\hline

$A_{N-1}$     & A    & $SU(N)$                 & $\frac{SL(N,C)}{SU(N)}$ & 2 & 0 & 0 & C$^+_{2,0,0}$ & G$^0_{2,0,0}$ & ${\rm T}^-_{2,0,0}$ \\ 
$A_{N-1}$     & AI   & $\frac{SU(N)}{SO(N)}$   & $\frac{SL(N,R)}{SO(N)}$ & 1 & 0 & 0 & C$^+_{1,0,0}$ & G$^0_{1,0,0}$ & ${\rm T}^-_{1,0,0}$\\ 
$A_{N-1}$     & AII  & $\frac{SU(2N)}{USp(2N)}$ & $\frac{SU^*(2N)}{USp(2N)}$ & 4 & 0 & 0 & C$^+_{4,0,0}$&G$^0_{4,0,0}$& ${\rm T}^-_{4,0,0}$\\
\hskip-2mm $\begin{array}{l} BC_q\ {\scriptstyle (p>q)} \\ C_q\  {\scriptstyle (p=q)} \end{array}$ 
                          & AIII & $\frac{SU(p+q)}{SU(p)\times SU(q)\times U(1)}$ & $\frac{SU(p,q)}{SU(p)\times SU(q)\times U(1)}$ & 2 & 1 & $2\nu $  & \hskip-2mm $\begin{array}{l}  \\ {\rm S}^+_{2,1,0}\end{array}$  & 
 $\chi^0_{2,1,2\nu }$ & 
 \hskip-2mm $\begin{array}{l}  \\ {\rm T}^-_{2,1,0} \end{array}$  \\
\hline

$B_N$          & B    & $SO(2N+1) $               & $\frac{SO(2N+1,C)}{SO(2N+1)}$ & 2 & 0 & 2  &   & P$^0_{2,0,2}$ &  \\ 

\hline

$C_N$ & C    & $USp(2N)$                               &$\frac{Sp(2N,C)}{USp(2N)}$ & 2 & 2 & 0 & B$^+_{2,2,0}$& B$^0_{2,2,0}$ & ${\rm T}^-_{2,2,0}$\\
$C_N$ & CI   & $\frac{USp(2N)}{SU(N)\times U(1)}$      & $\frac{Sp(2N,R)}{SU(N)\times U(1)}$& 1 & 1 & 0  & B$^+_{1,1,0}$ & B$^0_{1,1,0}$  & T$^-_{1,1,0}$  \\
\hskip-2mm $\begin{array}{l} BC_q\  {\scriptstyle (p>q)} \\  C_q\  {\scriptstyle (p=q)} \end{array}$ 
                  & CII & $\frac{USp(2p+2q)}{USp(2p)\times USp(2q)}$ & $\frac{USp(2p,2q)}{USp(2p)\times USp(2q)}$& 4 & 3 & $4\nu $ &   & $\chi^0_{4,3,4\nu }$ & \hskip-2mm $\begin{array}{l} \\ {\rm T}^-_{4,3,0}\end{array}$   \\
\hline

$D_N$    & D          &   $SO(2N)$                            & $\frac{SO(2N,C)}{SO(2N)}$ & 2 & 0 & 0 & B$^+_{2,0,0}$  &  B$^0_{2,0,0}$  & ${\rm T}^-_{2,0,0}$\\
$C_N$    & DIII  & $\frac{SO(4N)}{SU(2N)\times U(1)}$     & $\frac{SO^*(4N)}{SU(2N)\times U(1)}$ & 4 & 1 & 0 & B$^+_{4,1,0}$ & B$^0_{4,1,0}$ & T$^-_{4,1,0}$  \\
$BC_N$   & DIII  & $\frac{SO(4N+2)}{SU(2N+1)\times U(1)}$ & $\frac{SO^*(4N+2)}{SU(2N+1)\times U(1)}$& 4 & 1 & 4 &  & P$^0_{4,1,4}$ &   \\
\hline

\hskip-2mm $\begin{array}{l} B_q\ {\scriptstyle (p>q)}\\ D_q\ {\scriptstyle (p=q)} \end{array}$
            & BDI & $\frac{SO(p+q)}{SO(p)\times SO(q)}$   & $\frac{SO(p,q)}{SO(p)\times SO(q)}$ & 1 & 0 & $\nu $ &   & $\chi^0_{1,0,\nu }$ & $\begin{array}{l} \\ {\rm T}^-_{1,0,0} \end{array}$\\

\hline
\end{tabular}
\end{table}
\begin{table} 
\begin{center}
\setlength\tabcolsep{0.15cm}
\caption{
The correspondence between random matrix ensembles and symmetric 
spaces\label{tab2}} 
\begin{tabular}{|l|l|}
\hline
{\bf Random Matrix Theories (RMT)}           &  {\bf Symmetric Spaces (SS)} \\
\hline
\hline
circular or scattering ensembles           & positive curvature spaces      \\
\hline
gaussian or hamiltonian ensembles           & zero curvature spaces      \\
\hline
transfer matrix ensembles           & negative curvature spaces      \\
\hline
\hline
random matrix eigenvalues  & radial coordinates \\
\hline
probability distribution of eigenvalues  & Jacobian of transformation to radial
coordinates \\
\hline
Fokker--Planck equation        & radial Laplace--Beltrami equation         \\
\hline
Coulomb gas analogy              & Brownian motion on the symmetric space  \\
\hline
\hline
ensemble indices  &  root multiplicities \\
\hline
Dyson index $\beta $ &  multiplicity of ordinary roots ($\beta =m_o$)\\
\hline
boundary index $\alpha =\beta (\nu+1)-1$ & multiplicity of short and long roots
($\alpha=m_s+m_l$)\\
\hline
\hline
translationally invariant ensembles & SS with root lattice of type $A_n$ \\
\hline
boundary matrix ensembles  &  SS with root lattices of type $B_n$, $C_n$, $D_n$ or $BC_n$\\
\hline
\hline
pair interaction between eigenvalues & ordinary roots \\
\hline
\end{tabular}
\end{center}
\end{table}
\smallskip

\section{Solution of the DMPK equation}
\label{sec-dmpk}
\setcounter{equation}{0}

\begin{figure}[tb]
\centerline{
\psfig{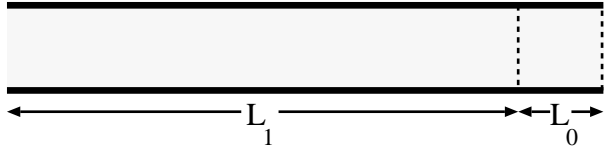}
}%
\medskip
\caption[]{
Disordered wire of length $L_{1}$ to which a segment of length $L_{0}$ is
added. This scaling operation leads to a Brownian motion of the transmission
eigenvalues determining the conductance. Taken from ref. \cite{Beenakker}. 
Used with permission.}
\label{fig_segment}
\end{figure}

One of the main problems in the RMT description of quantum wires is
determining the probability distribution of the $\{\lambda_i\}$
variables appearing in the parametrization of the transfer matrix,
eq.~(\ref{eq:Mpara}). This gives access to the transmission eigenvalues
and through the Landauer--Lee--Fisher formula, the main observable,
the conductance $G$ as a function of the length $L$ of the quantum
wire. To this end, Dorokhov \cite{Dorokhov} and Mello, Pereyra and
Kumar \cite{MPK} derived a scaling equation which expresses the
dependence of the probability distribution $P(\{\lambda_i\},L)$ of
$\{\lambda_i\}$ as a function of $L$.  This was done by considering
the change in $P(\{\lambda_i\},L)$ after addition of a thin slice $L_0$
to the wire under certain assumptions (cf. Fig.\ref{fig_segment}). 

The resulting equation reads

\begin{equation}
\frac{\partial P}{\partial s}=DP
\label{DMPK}
\end{equation}

where $s$ is the wire length measured in units of the mean free
path $l$: $s\equiv L/l$, and $D$ can be written in terms of the
$\{\lambda_i\}$ as follows:

\begin{equation}
D=\frac{2}{\gamma}\sum_{i=1}^{N}
\frac{\partial}{\partial\lambda_{i}}\lambda_{i}(1+\lambda_{i})
J_\beta(\lambda)\frac{\partial}{\partial\lambda_{i}}J_\beta(\lambda)^{-1},
\label{DMPK2}
\end{equation}

with $\gamma \equiv \beta N+2-\beta$.  $\beta $ is the symmetry index
of the ensemble of scattering matrices, in analogy with the
well--known Wigner--Dyson classification, and $J_\beta(\lambda )\equiv
J(\{\lambda_{i}\})$ is given by

\begin{equation}
J_\beta(\lambda )=\prod_{i<j}|\lambda_{j}-\lambda_{i}|^{\beta}
\label{jacobian}
\end{equation}

The solution of this equation was not at all immediate. What's more,
the appearance of $J_\beta(\{\lambda_{i}\})$ in the DMPK equation is
due to the fact that the authors tried to mimick the Wigner--Dyson
ensembles. However, this is quite misleading from the point of view of
symmetric spaces, as the Wigner--Dyson ensembles have nothing to do
with the transformation from the transfer matrix ensembles to the
space of the $\{\lambda_i\}$. As we will see, the final solution of
this equation, which is exact for $\beta=2$ and approximate for
$\beta=1,\ 4$, relies on the relationship of transfer matrices
with symmetric spaces of negative curvature.

The exact solution of the DMPK equation in the $\beta=2$ case was
first obtained in a remarkable paper \cite{BeeRejaei} by Beenakker and
Rejaei. Here we review their derivation in a slightly different
language, trying to stress the symmetric space origin of their result.

The starting point is a mapping of the coordinates $\lambda_i$
onto the radial coordinates $x_i$ on the symmetric space: 

\beq
\lambda_i \equiv {\rm sinh}^2x_i
\eeq

The DMPK equation can then be rewritten as

\begin{equation}
\frac{\partial P}{\partial s}=
\frac{1}{2\gamma}\, J(x)\Delta_B'J^{-1}(x)P
\label{DMPKbis2}
\end{equation}

where 

\begin{equation}
J(\{x_i\})=\prod_{i<j}|\sinh^{2}x_i-\sinh^{2}x_j|^\beta
\prod_{k}|\sinh 2x_{k}|
\label{nn1}
\end{equation}

is the Jacobian and $\Delta_B'$ is the radial part of the
Laplace--Beltrami operator on the underlying symmetric space! As a
consequence, we can identify the differential equation with the
equation for free diffusion on the symmetric space as a function of
dimensionless "time" $s=L/l$ where $l$ is the mean free path related
to diffusion in the quantum wire.

At this point one can follow two roads. On the first one, which was
the one followed by Beenakker and Rejaei, one maps the DMPK equation onto a
Schr\"odinger equation in imaginary time.  This requires the
substitution

\beq
\label{eq:subst}
P(\{x_i\},s) = J^{\frac{1}{2}}(\{x_i\}) \Psi(\{x_i\},s)
\eeq

A straightforward calculation shows that the DMPK equation then takes
the form 

\beq
\label{eq:DMPKSchr}
-\frac{\partial \Psi}{\partial s} = ({\cal H}-U)\Psi
\eeq

where $U$ is a constant and ${\cal H}$ is a Hamiltonian of the form

\beq
\label{eq:decoup}
{\cal H} = -\frac{1}{2\gamma}\sum_i \left( \frac{\partial^2}{\partial x_i^2}
+ {\rm sinh}^{-2}(2x_i) \right) + \frac{\beta (\beta -2)}{2\gamma}\sum_{i<j}
\frac{{\rm sinh}^2(2x_i) + {\rm sinh}^2(2x_j)}{({\rm cosh}(2x_i) -
{\rm cosh}(2x_j))^2}
\eeq

At this point the main goal has already been reached: it is easy to see
that if $\beta =2$ the equation decouples:

\begin{equation}
{\cal H}_{0}=-\frac{1}{2\gamma}\frac{\partial^{2}}
{\partial x^{2}}-\frac{1}{2\gamma\sinh^{2}2x}.
\label{hh1}
\end{equation}

The resulting equation was solved in \cite{BeeRejaei} using Greens functions.

The second approach to a solution of eq.(\ref{DMPKbis2}) also relies
on the underlying symmetric
space structure. Again, the starting point is the identification 
of the DMPK operator $D$ and the radial part
of the Laplace--Beltrami operator $\Delta'_B$ on the underlying symmetric
space: 

\begin{equation}
\frac{\partial P}{\partial s}=DP=\frac{1}{2\gamma}\, BP
\label{DMPKbis}
\end{equation}

where $\Delta_B'=J^{-1}BJ$.

As a consequence of this identification, if $\Phi_k(x)$ is an
eigenfunction of $\Delta_B'$ with eigenvalue $k^2$, then
$J(x)\Phi_k(x)$ will be an eigenfunction of the DMPK operator with
eigenvalue $k^2/(2\gamma)$.  Then the properties of the eigenfunctions
of the $\Delta_B'$ operator, which are zonal spherical functions, can
be used to derive the exact solution of the DMPK equation for
$\beta=2$ (\cite{M4}, see also \cite{M5}). This was done by
Caselle. As the details of this calculation are quite technical, we
will not review it here. The {\it exact} solution for the probability
distribution is in both cases

\begin{eqnarray}
P(\{x_{n}\},s)&=&C(s)\prod_{i<j}(\sinh^2x_j-\sinh^2x_i)
\prod_{k}(\sinh 2x_k)\nonumber\\
&&\mbox{}\times{\rm Det}\left[\int_{0}^{\infty}\!\!dk\,
\exp\left( -\frac{k^2s}{4N}\right) \tanh\left(\frac{\pi k}{2}\right) k^{2m-1}\,
{\rm P}_{\frac{1}{2}({\rm i}k-1)}(\cosh 2x_{n})\right]\nonumber \\
\label{final}
\end{eqnarray}

where ${\rm P}_{\nu}(z)$ denotes the Legendre functions of the first
kind. The second way of solving this equation constitutes a
non--trivial consistency check on the solution obtained by Beenakker
and Rejaei.

The power of the description in terms of symmetric spaces becomes
evident in the $\beta =1$ and $\beta =4$ cases, in which the
interaction between the eigenvalues does not vanish and the first
approach discussed above does not apply. On the contrary, the
description in terms of zonal spherical functions also holds in these
two cases. Even though for $\beta\not=2$ one does not know the
explicit form of the zonal spherical functions, one can use an
asymptotic expansion due to Harish--Chandra to get asymptotic
solutions. The asymptotic results derived by Caselle can be found in
\cite{MCDMPK}.

\section{Relation to Calogero--Sutherland models}
\label{sec-cs}
\setcounter{equation}{0}

An important role in our analysis is
played by the class of integrable models known as Calogero--Sutherland
(CS) models, which turn out to be deeply related to the theory of
symmetric spaces.  These models describe $n$ particles in one
dimension, identified by their coordinates $q^1,...,q^n$ and
interacting (at least in the simplest version of the models) through a
pair potential $v(q^i-q^j)$. The Hamiltonian of such a system 
is given by

\bea
\label{eq:calH}
{\cal H}=\frac{1}{2}\sum_{i=1}^n p_i^2 + \sum_{\alpha \in R^+} 
g_\alpha^2\, v(q^\alpha)\nonumber \\
p_i=-i\frac{\partial }{\partial q^i},\ \ \ \ 
q^\alpha = q\cdot \alpha =\sum_{i=1}^n q^i\alpha_i\\ \nonumber 
\eea

where the coordinate $q$ is $q=(q^1,...,q^n)$, $p_1,...,p_n$ are the
particle momenta, and the particle mass is set to unity.  In
eq.~(\ref{eq:calH}) $R^+$ is the subsystem of positive roots of the
root system $R=\{\alpha^1,...,\alpha^{\nu }\}$ related to a specific
simple Lie algebra or symmetric space, and $n$ is the dimension of the
maximal abelian subalgebra ${\bf H_0}$. The components of the positive
root $\alpha = \alpha^k\in R^+$ are $\alpha^k_1,...,\alpha^k_n$. The
number of positive roots is $\nu/2$, where $\nu $ is the total number
of roots. In general, the coupling constants $g_\alpha $ are
the same for equivalent roots, namely those that are connected with
each other by transformations of the Weyl group $W$ of the root system.
    
Several realizations of the potential $v(q^\alpha)$ have been studied
in the literature (for detailed discussions see the review by
Olshanetsky and Perelomov, ref.~\cite{OlshPere}). We will be concerned
with the following types of potentials:

\bea
\label{eq:I-V}
v_I(\xi )&=&\xi^{-2} \nonumber \\
v_{II}(\xi )&=&{\rm sinh}^{-2}\xi  \nonumber \\
v_{III}(\xi )&=&{\rm sin}^{-2}\xi  \\ \nonumber
\eea

{\bf Example:} The CS model corresponding to a $C_n$ root lattice is

 \beq
\label{CS3}
{\cal H}=-\frac{1}{2}\sum_{i=1}^{n}\frac{\partial^2}{\partial (q^i)^2}+
\sum_i\frac{g_l^2}{\sinh^2(2q^i)}
+\sum_{i<j}\left(\frac{g_o^2}{\sinh^2(q^i-q^j)}
+\frac{g_o^2}{\sinh^2(q^i+q^j)}\right)
\eeq

We remind the reader that the $C_n$ root system consists of root
vectors $\{\pm 2e_i,\pm e_i \pm e_j, i\neq j\}$.  The arguments of the
$\sinh $--function are $q^\alpha_k=q\cdot \alpha^k=(q^1,...,q^n)\cdot
(e_i\pm e_j)=q^i\pm q^j$ ($i<j$) where $\alpha^k $ is an ordinary
positive root of the root lattice $R$, and $(q^1,...,q^n)\cdot (\pm
2e_i)=\pm 2q^i$ for the long roots.  Note the different coupling
constants for the ordinary and long roots.

Under rather general conditions (see \cite{OlshPere} for a detailed
discussion) these models are completely integrable, in the sense that
they possess $n$ commuting integrals of motion. As we will now see,
the Calogero--Sutherland Hamiltonians can then be mapped onto the
radial parts of the Laplace--Beltrami operators on the relevant
symmetric spaces. The mapping is
 
\beq
\label{eq:H-Delta}
{\cal H}=J^{\frac{1}{2}}(q)\frac{1}{2}(\Delta_B'\pm\rho^2)J^{-\frac{1}{2}}(q)
\ \ \ \ \ \ \ (+\ \ {\rm for\ \ II},\ \ -\ \ {\rm for\ \ III},\ \ \rho =0\ \ 
{\rm for\ \ I})
\eeq

where $\rho $ is the vector defined in (\ref{eq:rho}) and $J(q)$ is
the Jacobian for transformation to radial coordinates on the SS:

\beq
\label{eq:xiJ}
\begin{array}{c}        \\
                J(q) \\
                        \\ \end{array}=
\begin{array}{ll} \prod_{\alpha \in R^+} [\,q^\alpha\,]^{m_\alpha } & {\rm I} \\
  \prod_{\alpha \in R^+} [\,{\rm sinh} (q^\alpha )\,]^{m_\alpha }  & {\rm II} \\
  \prod_{\alpha \in R^+} [\,{\rm sin} (q^\alpha )\,]^{m_\alpha }  & {\rm III} \\
\end{array}
\eeq

Olshanetsky and Perelomov proved that this happens {\it if and only
if} the coupling constants $g_\alpha $ in ${\cal H}$ take the
following {\it root values}

\beq
\label{eq:rootvalues_gen}
g_\alpha^2=\frac{m_\alpha (m_\alpha +2m_{2\alpha}-2)|\alpha |^2}{8}
\eeq

where $m_\alpha $ is the multiplicity of the root $\alpha $ and
$|\alpha |$ is its length.

A number of results can be obtained for the corresponding quantum
systems merely by using the theory of symmetric spaces. Due to
equation (\ref{eq:H-Delta}), once we know the zonal spherical
functions, we can solve the Schr\"odinger equation.  A detailed
collection of results pertaining to spectra, wave functions, and
integral representations of wave functions can be found in the
original article \cite{OlshPere}.

The equation (\ref{eq:decoup}) in the solution of the DMPK equation
discussed in the previous section can be recast in a slightly
different form, thus completing the chain of identifications
\vskip2mm

\centerline{\rm DMPK equation --- symmetric space ---
Calogero--Sutherland model} 

By using simple identities for hyperbolic functions, the Hamiltonian
in (\ref{eq:decoup}) becomes~\cite{MCDMPK}

\beq
\label{eq:HC_n}
\gamma{\cal H} = \sum_i \left( -\frac{1}{2}\frac{\partial^2}{\partial x_i^2}
+ \frac{g_l^2}{{\rm sinh}^2(2x_i)} \right) + \sum_{i<j} \left( 
\frac{g_o^2}
{{\rm sinh}^2(x_i-x_j)} + \frac{g_o^2}{{\rm sinh}^2(x_i+x_j)}\right) +c
\eeq

where $g_l^2\equiv -1/2$, $g_o^2\equiv\beta (\beta -2)/4$ and $c$ is
an irrelevant constant.  This Hamiltonian, apart from an overall
factor $1/\gamma $ and the constant $c$, exactly coincides with the
Calogero--Sutherland Hamiltonian (\ref{CS3}) corresponding to a root
lattice $R=\{\pm 2x_i,\pm x_i \pm x_j,i\neq j\}$ of type $C_n$ with
root multiplicities $m_o=\beta $, $m_l=1$.  The values of the coupling
constants $g_o$, $g_l$ are exactly the root values given in
eq.~(\ref{eq:rootvalues_gen}) for which the transformation from ${\cal H}$
into $\Delta_B'$ is possible.

\section{Concluding remarks}

In these lectures we have given an elementary introduction to random
matrix theory, as well as discussed some basic concepts in the theory
of symmetric spaces. The scope of these discussions were to show,
through numerous examples, that hermitean random matrix ensembles and
the symmetric coset spaces based on simple Lie groups and classified
by Cartan, are the same mathematical objects.  By making this
identification, new results can be obtained in random matrix theory by
applying what is known in mathematics about such manifolds. We gave a
few examples of this in the last lecture.

In the first lecture we discussed a number of systems that are
successfully described by random matrix theories. (Due to the limited
expertise of the author, only some major applications in physics were
discussed, however, there are also applications for instance in
mathematics and biology.)  We saw that random matrix theory gives a
statistical description of complex eigenvalue spectra of quantum
operators in chaotic or many--body systems. By taking the double
scaling limit in which the spectrum becomes dense and the eigenvalues
are simultaneously rescaled on the scale of the average level spacing,
one can separate out the universal part of the spectral behavior from
the system--dependent one. We briefly discussed some typical spectral
observables obtainable from the correlation functions. A large number
of systems (some of which were discussed in the introduction) exhibit
the generic spectral features arising in random matrix theory. 

One such system is represented by quarks in a random gauge field
background. We illustrated this by showing a few graphs of computer
simulations of QCD--like gauge theories on a space--time lattice. In
these the spectrum of the Dirac operator was studied and compared to
analytical predictions from random matrix theory. Of course, there is
a wealth of experimental and numerical results also on the spectra of
nuclei, atoms, molecules, elastomechanical systems, microwave
cavities, mesoscopic systems, etc., that we have not even mentioned
and that could be given as examples of the type of spectral behavior
that is predicted by random matrix theory. It is important to note
that RMT behavior in spectra can also be partial, depending on
circumstances not always fully understood (see \cite{GMW} for a
discussion). 

The typical eigenvalue repulsion in chaotic systems described by RMT
can be traced to the geometrical properties of the corresponding
symmetric space manifold. In particular we discussed how curvature,
type of roots, and root multiplicities of the restricted root lattice
on the symmetric manifold determine the exact form of these
correlations. This follows from the general theory of coordinate
systems on symmetric spaces and from the identification of the radial
coordinates on the symmetric space with the set of random matrix
eigenvalues. We made a number of further identifications between
matrix ensembles and symmetric spaces and summarized the results in 
Table~\ref{tab2}.

We also explained how symmetric spaces corresponding to compact
symmetric subgroups are classified. All possible involutive
automorphisms of the compact real form of an algebra leads, in a
natural way, to a classification based on root lattices. This defines
the universality classes of the corresponding random matrix ensembles,
and leads to a new scheme of classification dictated by the strict
properties of root systems belonging to simple complex Lie algebras.

The differential equation for quantum wires called the DMPK equation
describes the evolution with increasing wire length of the joint
probability distribution of a set of parameters simply related to the
transmission eigenvalues. In our discussion we concluded that it
essentially corresponds to the equation for free diffusion on the
symmetric space, and we discussed its solution in terms of zonal
spherical functions for all three values of the Dyson index. The zonal
spherical functions are known in the theory of symmetric spaces, where
they play the role of eigenfunctions for the radial part of the
Laplace--Beltrami operator on the symmetric manifold.

Lastly, we discussed work by Olshanetsky and Perelomov on the
integrability of a class of one--dimensional Calogero--Sutherland
models. Such models become integrable at certain ``root values'' of
the coupling constants, and the integrals of motion are then given by
the Casimir operators related to the Lie algebra or symmetric space
underlying the model. This leads to a set of exact results concerning
spectra and wave functions of quantum systems.

It is certainly of interest to further investigate the possibility of
applying known results on symmetric spaces to the corresponding
ensembles of random matrices. An important research direction might be
trying to fit also non--hermitean ensembles into a similar setting.
This would bring us outside Cartan's scheme. In spite of the recent
activity in the field of non--hermitean random matrix ensembles, the
author is not aware of any research effort in this direction.  There
are also various types of extensions of Calogero--Sutherland models
that could be explored in this spirit.

\end{document}